%% file: bsep_paper.tex
\documentclass[journal=jacsat,manuscript=article,layout=twocolumn]{achemso}
\usepackage[version=3]{mhchem} 
\usepackage{amssymb,color,amsmath,csvsimple,tabularx}
\usepackage[utf8]{inputenc}
\usepackage{subfiles}
\usepackage{numprint}
\usepackage{hyperref}
\usepackage{booktabs}
\usepackage{multirow}
\usepackage{makecell}
\usepackage{colortbl}
\usepackage{pdflscape}
\usepackage{afterpage}
\usepackage{capt-of}

\author{Kevin S. McLoughlin}
\affiliation{Lawrence Livermore National Laboratory}
\email{mcloughlin2@llnl.gov}
\author{Claire G. Jeong}
\affiliation{GlaxoSmithKline}
\author{Thomas D. Sweitzer}
\affiliation{GlaxoSmithKline}
\author{Amanda J. Minnich}
\affiliation{Lawrence Livermore National Laboratory}
\author{Margaret J. Tse}
\affiliation{GlaxoSmithKline}
\author{Brian J. Bennion}
\affiliation{Lawrence Livermore National Laboratory}
\author{Jonathan E. Allen}
\affiliation{Lawrence Livermore National Laboratory}
\author{Stacie Calad-Thomson}
\affiliation{GlaxoSmithKline}
\author{Thomas S. Rush}
\affiliation{GlaxoSmithKline}
\author{James M. Brase}
\affiliation{Lawrence Livermore National Laboratory}

\title{Machine Learning Models to Predict Inhibition of the Bile Salt Export Pump}

\begin{document}
\maketitle
\begin{abstract}
\subfile{sections/Abstract.tex}
\end{abstract}

\section{Introduction}
\subfile{sections/Introduction.tex}

\section{Methods}

\subfile{sections/Methods.tex}

\section{Results and Discussion}\label{sec:results}

\subfile{sections/Results.tex}

\section{Conclusions}
\subfile{sections/Conclusion.tex}

\section{Disclaimer}
This document was prepared as an account of work sponsored by an agency of the United States government. Neither the United States government nor Lawrence Livermore National Security, LLC, nor any of their employees makes any warranty, expressed or implied, or assumes any legal
liability or responsibility for the accuracy, completeness, or usefulness of any information, apparatus, product, or process disclosed, or represents that its use would not infringe privately owned rights. Reference herein to any specific commercial product, process, or service by trade name, trademark, manufacturer, or otherwise does not necessarily constitute or imply its endorsement, recommendation, or favoring by the United States government or Lawrence Livermore National Security, LLC. The views and opinions of authors expressed herein do not
necessarily state or reflect those of the United States government or Lawrence Livermore National Security, LLC, and shall not be used for advertising or product endorsement purposes. 

This work was performed under the auspices of the U.S. Department of Energy by Lawrence Livermore National Laboratory under contract DE-AC52-07NA27344.

\bibliography{bsep_paper}

\section{Supplemental Information}

\subfile{sections/Supplement.tex}

\subfile{sections/SuppTables.tex}

\end{document}

%% file: sections/Abstract.tex
Drug-induced liver injury (DILI) is the most common cause of acute liver failure and a frequent reason for withdrawal of candidate drugs during preclinical and clinical testing. An important type of DILI is cholestatic liver injury, caused by buildup of bile salts within hepatocytes; it is frequently associated with inhibition of bile salt transporters, such as the bile salt export pump (BSEP). Reliable in silico models to predict BSEP inhibition directly from chemical structures would significantly reduce costs during drug discovery and could help avoid injury to patients. Unfortunately, models published to date have been insufficiently accurate to encourage wide adoption. We report our development of classification and regression models for BSEP inhibition with substantially improved performance over previously published models. Our model development leveraged the ATOM Modeling PipeLine (AMPL) developed by the ATOM Consortium, which enabled us to train and evaluate thousands of candidate models. In the course of model development, we assessed a variety of schemes for chemical featurization, dataset partitioning and class labeling, and identified those producing models that generalized best to novel chemical entities. Our best performing classification model was a neural network with ROC AUC = 0.88 on our internal test dataset and 0.89 on an independent external compound set. Our best regression model, the first ever reported for predicting BSEP IC50s, yielded a test set $R^2 = 0.56$ and mean absolute error 0.37, corresponding to a mean 2.3-fold error in predicted IC50s, comparable to experimental variation. These models will thus be useful as inputs to mechanistic predictions of DILI and as part of computational pipelines for drug discovery.

%% file: sections/Introduction.tex
Drug induced liver injury (DILI) has been reported to be the leading cause of acute liver failure in the US\cite{schiodt_etiology_1999}.  Incidence of DILI frequently results in early termination of preclinical development and clinical trials for candidate drugs, and imposition of  black-box warnings or outright withdrawal of marketed drugs\cite{holt_mechanisms_2006,tujios_mechanisms_2011}. There are two main forms of DILI: the more severe hepatocellular form, associated with elevated alanine transferase; and the less severe cholestatic form, characterized by elevated alkaline phosphatase\cite{hussaini_idiosyncratic_2007}. Drugs can cause DILI via several different and overlapping mechanisms. In spite of considerable efforts, the complex, multifactorial nature of DILI renders its mechanistic understanding and prediction difficult. The most common mechanisms covered by in vitro assays include formation of reactive metabolites; inhibition of transporters involved in the export of bile acids from hepatocytes; mitochondrial toxicity; and other cellular toxicity pathways. One likely contributor to DILI is inhibition of the bile salt export pump (BSEP) transporter, leading to increase of intracellular bile salt concentrations to toxic levels\cite{dawson_vitro_2012,thompson_vitro_2012,shah_setting_2015}. 

BSEP, encoded by the gene ABCB11, belongs to a family of ATP-binding cassette (ABC) transporters that also includes P-glycoprotein (ABCB1) and the multidrug resistance proteins MRP2 (ABCC2), MRP3 (ABCC3) and MRP4 (ABCC4). These transporters mediate the excretion of individual bile constituents from hepatocytes into canaliculi or blood vessels and play key roles in bile formation and cholestasis. BSEP, mainly expressed in the liver, is localized in the cholesterol-rich canalicular membrane of hepatocytes with twelve transmembrane-spanning domains. Its function is to export unconjugated and conjugated bile acids and salts from hepatocytes into the bile, and to thereby maintain a low intracellular concentration of bile salts. Particularly in humans, impairment of BSEP function can lead to disturbed bile salt excretion and to cholestasis. 

Inhibition of BSEP has been clearly linked to cholestatic liver injury in several studies\cite{dawson_vitro_2012,morgan_interference_2010}. While several transporters may be involved in cholestasis, drug-like compounds are more likely to inhibit BSEP than the other transporters. In a recent study of 635 currently and formerly marketed drugs\cite{morgan_multifactorial_2013}, 148 (23\%) inhibited BSEP function with IC50s less than 100 $\mu$M, and 78 (12\%) with IC50s less than 25 $\mu$M. By contrast, only 4\% and 7\% of compounds tested had IC50s less than 100 $\mu$M for MRP2 and MRP3, respectively. MRP4 inhibition was less common than BSEP inhibition but more common than MRP2 and MRP3, with 17\% of compounds measured having IC50s below 100 $\mu$M and 9\% with IC50s less than 25 $\mu$M. 

While in vitro assays for BSEP inhibition can be performed with moderately high throughput, they are currently too expensive to apply routinely to screen hundreds or thousands of candidate drugs, and the cost of synthesizing lead compounds precludes the use of these assays in the earliest stages of drug discovery. Therefore, in silico approaches to identify compounds as potential BSEP inhibitors would be much more cost effective.

Some groups have addressed the need for computational prediction of BSEP inhibition through mechanistic approaches, such as docking and molecular dynamics simulations\cite{jain_structure_2017}. Unfortunately, accurate mechanistic models of protein-ligand interactions require high resolution protein structures, which are not currently available for BSEP. While approximate structures have been constructed based on the homology of BSEP and P-glycoprotein, the predictive accuracies of models based on these structures lag behind the accuracy of ligand-based models. In addition, such models are complicated by the difficulty of simulating interactions with the surrounding cell membrane.

As a result, efforts to predict BSEP inhibition have focused on machine learning approaches, based on structural features of small molecule ligands. One of the earliest published models\cite{hirano_high-speed_2006} was a multiple linear regression model, which fit single-concentration percent inhibition values to a linear combination of “chemical fragmentation codes” (occurrence counts of particular structural motifs). The authors fit the model to data for 38 drugs and reported a coefficient of determination ($R^2$) of 0.952 across the training dataset. The authors did not perform any cross-validation to determine how well their model generalizes to compounds outside their training set. 

Warner et al.\cite{warner_mitigating_2012} built several classification models based on descriptor features calculated by AstraZeneca in-house software for 624 compounds, assigned randomly to training (70\%) and test sets (30\%). Compounds were categorized as BSEP inhibitors if their IC50s were below 300 $\mu$M. Their best model was a support vector machine which yielded 87\% accuracy, precision 0.85 and recall 0.90. While their model depends on in-house descriptor calculations and was not published, their dataset was published as supporting information.

Pedersen et al.\cite{pedersen_early_2013} developed partial least-squares (PLS) classification models using descriptors calculated by a combination of in-house and commercial software (Dragon 6, ADMETPredictor). Their dataset consisted of 249 compounds, randomly split into training (65\%) and test sets (35\%). Compounds were classified based on their percent inhibition of BSEP function at 50 $\mu$M concentration: as weak inhibitors if they reduced taurocholic acid (TA) transport to between 50\% and 72.5\% of control values, and as strong inhibitors if they reduced it to 50\% or less. The PLS model for discriminating strong inhibitors from non-inhibitors uses a combination of 15 descriptors; the authors report 89\% accuracy on the test set, with precision 0.84 and recall 0.76. Weak inhibitors were excluded from the test set for this model. A model for discriminating all inhibitors (strong and weak) from non-inhibitors performed less well, with 73\% accuracy, precision 0.65 and recall 0.63. The models are not publicly available, but the dataset is included within the article.

Montanari et al.\cite{montanari_flagging_2016} built a random forest model using IC50 data for 838 compounds contributed by AstraZeneca to the IMI eTox project (\url{http://etoxproject.eu}). The feature set consisted of 78 descriptors calculated by the commercial software MOE\cite{ccg_molecular_2018}. According to the authors, the descriptors were chosen based on interpretability, rather than information content or correlation with BSEP inhibition. Compounds were labeled as inhibitors if the IC50 was less than or equal to 10 $\mu$M, and as noninhibitors if the IC50 was greater than 50 $\mu$M. Notably, weak inhibitors with IC50s between 10 and 50 $\mu$M were excluded from the training and testing datasets. Compounds were assigned randomly to training (80\%) and test sets (20\%). In addition, the authors compiled an external validation dataset of 156 compounds, taken from the Hirano et al. study and from a dataset published by Morgan et al.\cite{morgan_interference_2010}. The IC50 thresholds used to label compounds as inhibitors or non-inhibitors in the external datasets differed somewhat from those used for the training and testing data, apparently due to differences in the assays used to generate the data. To select the best number of trees for random forest models, the authors used 5-fold and 10-fold cross-validation. The best model yielded 80\% accuracy on the test set, with precision 0.70 and recall 0.67. On the external test set, accuracy was 88\%, with precision 0.77 and recall 0.77. To our knowledge, Montanari et al. were the first to publish their model; it is available as a KNIME workflow that requires the user to have a license for MOE to compute the necessary descriptors.

Although the published models show promising performance when tested on existing pharmaceutical compounds, it remains uncertain how well these models generalize to compounds with novel chemistry. One area of concern is the method used to partition compounds into training and test sets. All three of the classification models discussed above were trained using random splitting to select a test set. As pointed out by Rohrer\cite{rohrer_maximum_2009}, Wallach\cite{wallach_most_2018} and others, random selection leads to biased test sets in which many test compounds are structurally similar to compounds in the training set. Model selection based on performance on such test sets tends to favor models that “memorize” the training data, rather than those which learn to generalize over common chemical features. Several dataset splitting strategies have been proposed to improve model generalizability\cite{wallach_most_2018,butina_unsupervised_1999,ramsundar_deep_2019}, but have yet to be employed to train predictive models of BSEP inhibition.

Another concern is the use of dual IC50 or percent inhibition thresholds to classify compounds as BSEP inhibitors or non-inhibitors\cite{montanari_flagging_2016}. Excluding weak inhibitors from the training and testing datasets simplifies the classification task, by removing compounds with intermediate IC50 values that are inherently most difficult to classify. When applied in a realistic drug discovery context, however, the model will have to predict inhibition for compounds whose true IC50s may fall in the intermediate range. Therefore, performance metrics reported for test sets with weak inhibitors excluded are likely to be inflated relative to what can be expected in a realistic drug discovery scenario.

Except for the model reported by Hirano et al.\cite{hirano_high-speed_2006}, which predicts percent inhibition of BSEP function at a fixed concentration, we are unaware of any published regression models for BSEP inhibition. Models that can predict an IC50 or the percent inhibition at a specified concentration are needed to provide input to quantitative systems models for hepatotoxicity such as DILIsym\cite{watkins_dilisim_2019} and to computational drug design frameworks. However, regression models are difficult to fit to IC50 data because many data values are censored (e.g., the IC50 is reported as “greater than 100 $\mu$M” if the compound was tested at a maximum concentration of 100 $\mu$M, and the transporter activity was not at least 50\% reduced at that concentration).

Our hypothesis for the research reported here was that, by applying rigorous machine learning techniques and extensive hyperparameter searches, one could create a set of predictive models for BSEP inhibition that would better generalize to molecules with novel chemistry. We describe here the process we devised to train and evaluate models in order to avoid the shortcomings of previous work. In particular, we will show that the method used for splitting data into training, validation and test sets has a large effect on measured performance and generalization power. We compare the performance of different types of models (i.e., neural networks and random forests) using different feature sets (ECFP fingerprints and chemical descriptors). We also demonstrate the inflation of performance metrics resulting from use of dual-threshold classification schemes and show how a single threshold leads to more realistic estimates of performance on novel compounds. We present results from training both classification and regression models, going beyond previous work that addressed the classification problem only. 

Our research serves as an application case study for the open source ATOM Modeling PipeLine (AMPL)\cite{minnich_ampl_2019}, developed by our team at the Accelerating Therapeutics for Opportunities in Medicine (ATOM) Consortium. In order to facilitate reproduction of our research, we have released example models for BSEP inhibition based on open data and descriptor calculations, which are available on the AMPL GitHub repository (\url{https://github.com/ATOMconsortium/AMPL}). We discuss the performance of these models and compare them to models trained using mixed open and proprietary data.

%% file: sections/Methods.tex
\subsection{Data curation}
We constructed a combined training and testing data set using data from two sources: A proprietary BSEP assay data set donated to the ATOM Consortium by GSK, containing data from 601  compounds; and a published data set included as supplemental data by Morgan et al.\cite{morgan_multifactorial_2013}, containing data for 634 compounds. Both data sets were based on measurements of [3H]-taurocholate transport in inverted membrane vesicles, following similar protocols. For the Morgan et al. data set, we used the ChEMBL 25 database\cite{gaulton_chembl_2017} to map the published compound names to ChEMBL compound IDs and SMILES strings; we found unambiguous matches for 631 compounds. For both data sets, we used the RDKit\cite{landrum_rdkit_2018} and MolVS\cite{swain_molvs:_2016} Python packages to standardize SMILES strings and remove salts. We excluded organometallic compounds, compounds with molecular weight greater than 2000 or less than 100, and mixtures of molecular species (e.g. aminophylline) from both data sets. For modeling and analysis, IC50 values were converted to pIC50 values, where pIC50 = $-\log_{10} (IC50)$ when IC50 is expressed as molar concentration.

\begin{figure}
    \centering
    \includegraphics[width=.5\textwidth]{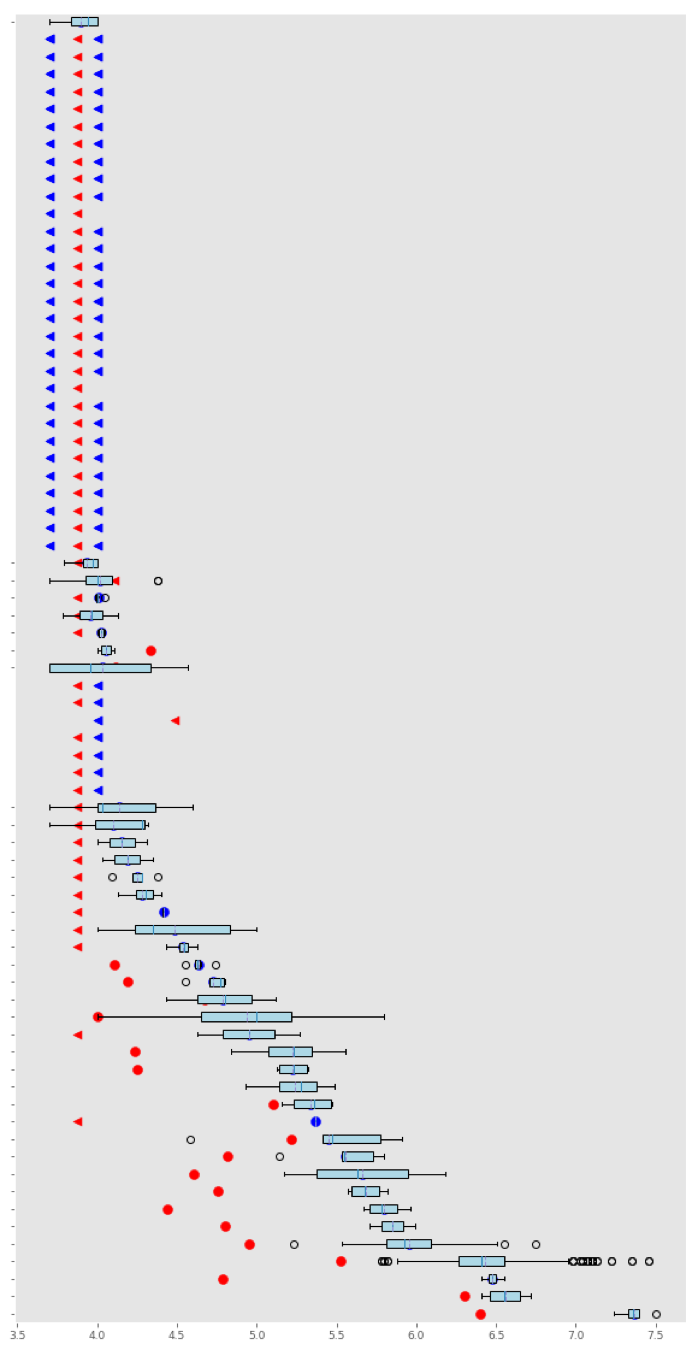}
    \caption{Measured pIC50 values for compounds appearing in both the Morgan et al. (red points) and GSK (blue) data sets. Compounds with replicate uncensored values in the GSK data set are represented by box plots showing the interquartile range of replicate measurements. Triangles represent left-censored pIC50 values, corresponding to the maximum concentrations tested (133 $\mu$M in the Morgan et al. data set, 100 or 200 $\mu$M in the GSK data set).}
    \label{fig:gsk_vs_public}
\end{figure}

The resulting data sets shared 75 compound structures in common. We compared the measured pIC50 values for these compounds between the two data sets, as shown in Figure \ref{fig:gsk_vs_public}. Although the GSK assay tended to report larger pIC50 values for compounds with measurable BSEP inhibition, the two assays usually agreed as to whether compounds should be categorized as inhibitors or non-inhibitors. We chose a threshold pIC50 = 4, corresponding to an IC50 of 100 $\mu$M, based on the maximum concentrations tested for the two data sets (100 or 200 $\mu$M for the GSK data, and 133 $\mu$M for the Morgan et al. data set); compounds with reported pIC50 $> 4$ were categorized as inhibitors. For the GSK data, where most compounds had replicate measurements, compounds were categorized as inhibitors if a strict majority of pIC50 values were greater than 4. Based on these criteria, the reported values from GSK and Morgan et al. produced the same categorization for 62 out of 75 compounds. Nine compounds were classified as inhibitors according to the Morgan et al. data but not the GSK data; while four compounds were inhibitors according to the GSK data but not the Morgan et al. data. In all but one of the compounds where the data sets disagreed, the average reported pIC50 for the data set designating the compound as an inhibitor was between 4 and 5, suggesting that these were borderline cases. 

We concluded that the assay data from the two data sets were sufficiently comparable to justify combining them into one classification data set. To combine the data sets, we used a maximum likelihood procedure to estimate a mean pIC50 over replicate measurements, while taking censoring into account; when a compound was in both the GSK and Morgan et al. data sets, we treated the value from Morgan et al. as an additional replicate. Details about the estimation procedure are given in the supplemental information.

To compare model performance against the results of Montanari et al., we also constructed a dual-threshold training and testing data set from which compounds with IC50s between 10 and 50 $\mu$M (pIC50s between 4.3 and 5.0) were excluded. In this dataset, compounds were classified as inhibitors if their IC50 was less than 10 and as non-inhibitors if IC50 was greater than 50 $\mu$M.

To provide further assessment of model performance and generalization power, we compiled an external test data set from the data published by Warner et al.\cite{warner_mitigating_2012}. We used Adobe Acrobat DC to convert the PDF table to an Excel spreadsheet, manually edited the spreadsheet to align rows and columns consistently and used an in-house Python script to remove embedded newlines and blanks in SMILES strings and compound names. SMILES strings were canonicalized and salt groups removed as for our training dataset. We identified 142 compounds appearing both in the Warner et al. dataset and our combined training/testing dataset and used these to assess the comparability of IC50 measurements between the two. We found that IC50 measurements in the Warner et al. dataset tended to be smaller than corresponding measurements in our internal dataset, probably due to differences in assay protocols. Therefore, to create our external classification dataset, we used a lower IC50 threshold to label compounds as inhibitors. We chose the threshold by computing Cohen’s kappa statistic between our classifications and a set of trial classifications based on thresholds ranging from 15 to 130 and selecting the threshold that maximized kappa. The optimal threshold was 50 $\mu$M, which yielded a kappa value of 0.62. We constructed the external classification dataset using the remaining data, filtered to exclude compounds having Tanimoto similarity greater than 0.6 to any compound in our internal training and testing set (based on ECFP4 fingerprints calculated by RDKit). We also excluded compounds containing metals or with molecular weight less than 100. The final dataset contained classification data for 366 compounds. We did not create an external regression test set, due to the limited overlap between the IC50 ranges in our dataset (1 nM – 133 $\mu$M) and that of Warner et al. (10 – 1000 $\mu$M).

To generate the open-data classification model released with this paper, we prepared an additional training dataset by selecting records from the datasets published by Morgan et al. and Warner et al. For compounds present in both datasets, we selected the data from the Morgan et al. set, because the IC50 measurements in that set covered a wider range. As before, we labeled compounds as BSEP inhibitors if their IC50 was below 50 $\mu$M, for compounds 
derived from the Warner et al. dataset, or 100 $\mu$M, for compounds from the Morgan et al. dataset. The combined set of compounds contained a much smaller percentage of inhibitors (24\%) than the combined Morgan/GSK dataset (44\%). To improve the performance of open-data models, we created a class-balanced dataset containing all the inhibitors plus a random sample of non-inhibitors, totaling 660 compounds, 45\% of which were inhibitors. For external validation of open-data models, we used a subset of GSK compounds filtered to exclude those with Tanimoto similarity greater than 0.6 to any compound in the training set.

\subsection{Featurization}
SMILES strings from the combined dataset were mapped to model input features using three different featurization schemes. 
Chemical descriptors were computed using MOE using both 2D and 3D structures. We noted that many MOE descriptors were strongly correlated with each
other due to the fact that they scaled with molecular size, as measured by the total number of atoms \texttt{a\_count}.
To remove this dependency, we replaced all descriptors $d$ having Pearson correlation $r(d, \mathtt{a\_count}) > 0.5$ with scaled versions, 
computed as $d/\mathtt{a\_count}$. The resulting set of features was pruned further to eliminate descriptors that duplicated or were 
linear functions of other descriptors or had constant values.

An alternative set of chemical descriptors was computed using the open source package “Mordred”\cite{moriwaki_mordred:_2018}. The complete set of 2D and 3D descriptors was pruned to exclude those that were undefined or noncomputable for the majority of compounds in the ATOM database. No scaling or other transformations were applied to Mordred descriptors. The resulting feature set contained 1,555 descriptors. For the open-data model released with this paper, Mordred descriptors were used exclusively, in order to make the model usable without the need for commercial software.

Finally, extended connectivity fingerprint (ECFP4) bit vectors with length 1024 were computed using RDKit.

\subsection{Dataset partitioning}
The combined proprietary (GSK) and public (Morgan et al.) dataset was partitioned into training, validation and test sets containing 70\%, 15\% and 15\% of the compounds respectively. Three different partitioning strategies were compared: random assignment, the “scaffold” splitter implemented in the DeepChem package\cite{ramsundar_deep_2019}, and a modified version of the asymmetric validation embedding (AVE) debiasing splitter described by Wallach et al.\cite{wallach_most_2018}. The scaffold splitter aims to increase chemical dissimilarity between the training, validation and test sets by mapping compound structures to their Bemis-Murcko scaffolds and assigning compounds with the same scaffold to the same partitions, so that test compounds do not share scaffolds with training compounds. 

The AVE debiasing splitter aims to produce a split that provides an improved estimate of generalization performance. It uses a genetic algorithm to optimally partition compounds into training and test sets in order to minimize a quantity called the AVE bias. Partitions with large AVE bias favor models that memorize the training data, such as k-nearest neighbor models; conversely, minimizing this bias should reward models that generalize beyond the training data. The AVE bias is the sum of two components: one that measures the likelihood that an inhibitor in the test set is most chemically similar to an inhibitor in the training set, and another representing the likelihood that a non-inhibitor in the test set is most similar to a non-inhibitor in the training set. To compute similarities between training and test set compounds and identify the nearest neighbors for each test compound, we used a distance metric appropriate to the type of features used in the model: Tanimoto distance for ECFP features, and Euclidean distance for MOE or Mordred descriptors. Details of our modifications to the published AVE debiasing algorithm are given in the supplement. 
Note that in our implementation, the AVE debiasing split is applied twice, first to select a held-out test set, and again to choose 
compounds for the validation set. Because the validation set is used to select optimal model parameters (as described below), we hypothesized that 
models trained with the AVE debiasing split would perform better against external test compounds than models trained with other splitting algorithms.

\subsection{Model training}
Neural network and random forest models were trained and evaluated using a data-driven modeling pipeline, AMPL, developed by our group at the ATOM Consortium\cite{minnich_ampl_2019}. Neural networks consisted of one, two or three fully connected hidden layers, with varying numbers of rectified linear unit (ReLU) nodes per layer. During training and evaluation, 30\% of nodes were dropped out randomly to avoid overfitting. Random forest models were trained with varying numbers of decision trees, maximum depth, and maximum numbers of features used per split. The underlying models were implemented with the DeepChem and scikit-learn packages. Both classification and regression models were trained and tested. 

\begin{figure}
    \centering
    \includegraphics[width=.45\textwidth]{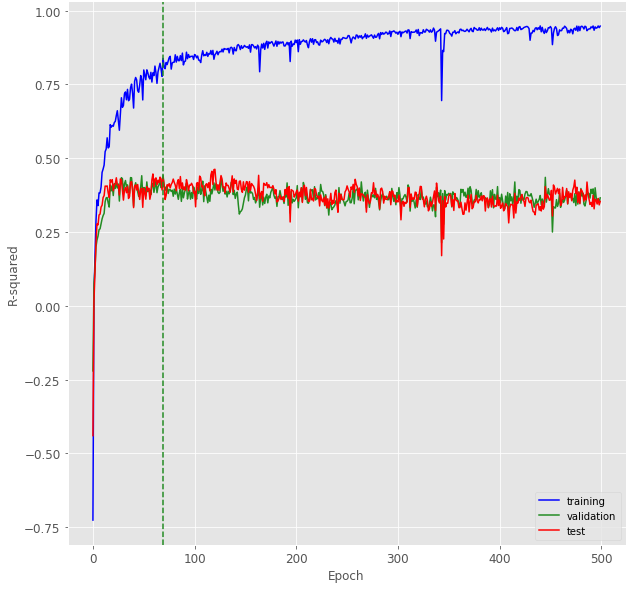}
    \caption{Illustration of early stopping procedure for neural network training. The graph shows the evolution of regression model performance ($R^2$) over the course of training for the training (blue), validation (green) and test sets (red). Neural network weights for the final model are taken from the end of the training epoch yielding the best $R^2$ for the validation set, indicated here by a vertical dashed line.}
    \label{fig:perf_vs_epoch}
\end{figure}

As an additional strategy to avoid overfitting neural network models, we implemented an early stopping procedure as follows: Models were evaluated after each training epoch, using the validation set to compute the area under the receiver operating characteristic curve (ROC AUC) for classification, or the coefficient of determination ($R^2$) for regression models. After training for a preset number of epochs (500), we noted the epoch yielding the maximum validation score and used the corresponding network weights as the parameters for the final model. The early stopping scheme is illustrated in Figure \ref{fig:perf_vs_epoch}. Note that, while the training set $R^2$ score continues to increase, the validation and test set scores reach a peak and then gradually decline, as the model becomes more overfitted to the training set data.

To optimize neural network performance, we ran hyperparameter searches for each splitting method and feature type, varying the numbers of hidden layers, number of nodes per layer, and learning rate. We ran similar hyperparameter searches for random forest models, varying the number of trees, maximum tree depth and maximum features per split. Model parameters and performance metrics were stored in a MongoDB database for subsequent analysis; over 6,600 classification models and 8,900 regression models were evaluated.

\subsection{Performance evaluation}
Following standard machine learning practices, we selected the best model parameters for each choice of model type, feature set and splitting strategy based on validation set performance metrics: ROC AUC for classification models and $R^2$ for regression models. We then evaluated a variety of performance metrics for the best models against the held-out test set and report these values here. In addition to the ROC AUC, the metrics reported for classification models include the following, defined in terms of the counts of true positives ($TP$), true negatives ($TN$), false positives ($FP$) and false negatives ($FN$):
\begin{itemize}
    \item Area under the precision vs recall curve (PRC AUC). 
    \item Precision or positive predictive value (PPV):
    \begin{equation*}
        \frac{TP}{TP+FP}
    \end{equation*}
    \item Negative Predictive Value (NPV):
    \begin{equation*}
        \frac{TN}{TN+FN}
    \end{equation*}
    \item Recall or sensitivity:
    \begin{equation*}
        \frac{TP}{TP+FN}
    \end{equation*}
    \item Accuracy:
    \begin{equation*}
        \frac{TP+TN}{TP+TN+FP+FN}
    \end{equation*}
    \item Matthews correlation coefficient (MCC):
    \begin{equation*}
        \frac{TP \cdot TN - FP \cdot FN}{\sqrt{(TP+FP)(TP+FN)(TN+FP)(TN+FN)}}
    \end{equation*}
\end{itemize}

For regression models, we report the following performance metrics, defined in terms of the true pIC50 values $y_i$, the predicted values $\hat{y}_i$ for each compound, and the mean true value $\bar{y}$:
\begin{itemize}
    \item Coefficient of determination ($R^2$):
    \begin{equation*}
        1 -\frac{\sum_{i=1}^{n} (y_i - \hat{y_i})^2}{\sum_{i=1}^{n} (y_i - \bar{y})^2}
    \end{equation*}
    \item Mean absolute error (MAE):
    \begin{equation*}
        \frac{1}{n} \sum_{i=1}^{n} |y_i - \hat{y_i}|
    \end{equation*}
    \item Root mean square error (RMSE):
    \begin{equation*}
        \sqrt{\frac{1}{n} \sum_{i=1}^{n} (y_i - \hat{y_i})^2}
    \end{equation*}
\end{itemize}

%% file: sections/Results.tex
\subsection{Performance of BSEP classification models}
\begin{figure*}[tp]
    \centering
    \includegraphics[width=.95\textwidth]{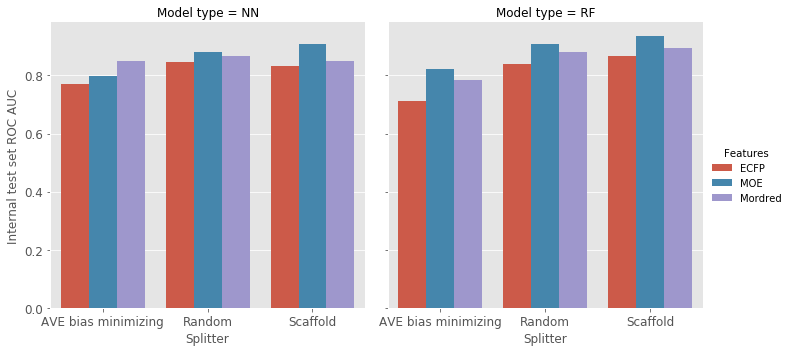}
    \caption{Performance of best classification models for each model type, feature type, and data set splitting method, computed as ROC AUC for test set predictions.}
    \label{fig:classif_perf}
\end{figure*}

Figure \ref{fig:classif_perf} summarizes the performance, measured as the test set ROC AUC, of the best classification model for each combination 
of model type (random forest or neural network), data set splitting method, and type of chemical features. Characteristics of the models and 
ROC AUC scores for each dataset partition are shown in Tables \ref{tab:nn_classif_perf} and \ref{tab:rf_classif_perf}. 
More detailed tables of model characteristics and performance metrics are given in Supplemental Tables S1 and S2. The figure makes it clear that the splitting algorithm has a marked effect on measured performance. Generally speaking, the AVE bias minimizing splitter provided the most stringent performance assessment, with AUC scores averaging about 0.1 units smaller than the corresponding scores for random and scaffold splits. Among the best models of a given type trained with the same splitter, those using MOE descriptor features almost always performed better; the exceptions were the neural network models trained with the AVE debiasing splitter, for which the highest ROC AUC score was achieved with Mordred descriptors.

\begin{figure*}[tp]
    \centering
    \includegraphics[width=.95\textwidth]{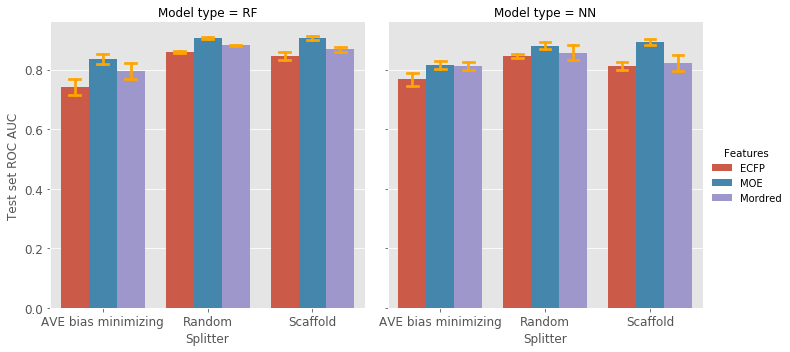}
    \caption{Variation of test set ROC AUC metrics across 10 identically trained classification models for each model type, feature type, and data set splitting method. Bars show mean values with error bars spanning $\pm$ 1 SD.}
    \label{fig:classif_perf_var}
\end{figure*}

\begin{table*}[tp]
  \centering
  \caption{Hyperparameters and ROC AUC scores for best neural network classification models. Test set ROC AUCs for best performing models within each splitter class are highlighted in yellow.}
  \resizebox{\textwidth}{!}{%
    \begin{tabular}{|l|lccc|cccc|}
    \toprule
    \multicolumn{1}{|c|}{\multirow{2}[4]{*}{\textbf{Splitter}}} & \multicolumn{1}{c}{\multirow{2}[4]{*}{\textbf{Features}}} & \multicolumn{1}{c}{\multirow{2}[4]{*}{\textbf{\# Hidden layers}}} & \multicolumn{1}{c}{\multirow{2}[4]{*}{\textbf{Hidden layer sizes}}} & \multicolumn{1}{c|}{\multirow{2}[4]{*}{\textbf{Training epochs}}} & \multicolumn{4}{c|}{\textbf{ROC AUC}} \\
\cmidrule{6-9}          &       &       &       &       & \multicolumn{1}{p{4.085em}}{Training set} & \multicolumn{1}{p{4.835em}}{Validation set} & \multicolumn{1}{p{3.25em}}{Test set} & \multicolumn{1}{p{3.585em}|}{External set} \\
    \midrule
    \multicolumn{1}{|l|}{\multirow{3}[2]{*}{AVE bias minimizing}} & ECFP  & 2     & 64,8  & 2     & 0.952 & 0.699 & 0.770 & 0.752 \\
          & MOE   & 2     & 12,7  & 155   & 0.990 & 0.866 & 0.799 & 0.836 \\
          & Mordred & 2     & 15,5  & 10    & 0.926 & 0.834 & \cellcolor[rgb]{ 1,  1,  0}0.849 & 0.856 \\
    \midrule
    \multirow{3}[2]{*}{Random} & ECFP  & 3     & 14,12,7 & 13    & 0.959 & 0.913 & 0.846 & 0.773 \\
          & MOE   & 3     & 15,6,3 & 74    & 0.967 & 0.911 & 0.882 & 0.883 \\
          & Mordred & 3     & 13,5,3 & 96    & 0.969 & 0.950 & 0.868 & 0.854 \\
    \midrule
    \multirow{3}[2]{*}{Scaffold} & ECFP  & 2     & 16,11 & 4     & 0.918 & 0.883 & 0.833 & 0.771 \\
          & MOE   & 3     & 32,16,4 & 128   & 0.969 & 0.905 & 0.907 & 0.816 \\
          & Mordred & 3     & 64,16,1 & 17    & 0.914 & 0.899 & 0.851 & 0.845 \\
    \bottomrule
    \end{tabular}%
  \label{tab:nn_classif_perf}%
}
\end{table*}%

\begin{table*}[tp]
  \centering
  \caption{Hyperparameters and ROC AUC scores for best random forest classification models. Test set ROC AUCs for best performing models within each splitter class are highlighted in yellow; the best overall ROC AUC is highlighted in orange.}
  \resizebox{\textwidth}{!}{%
    \begin{tabular}{|l|lccc|cccc|}
    \toprule
    \multicolumn{1}{|c|}{\multirow{2}[4]{*}{\textbf{Splitter}}} & \multicolumn{1}{c}{\multirow{2}[4]{*}{\textbf{Features}}} & \multirow{2}[4]{*}{\textbf{\# Trees}} & \multirow{2}[4]{*}{\textbf{Max depth}} & \multicolumn{1}{c|}{\multirow{2}[4]{*}{\textbf{Max features per split}}} & \multicolumn{4}{c|}{\textbf{ROC AUC}} \\
\cmidrule{6-9}          &       &       &       &       & \multicolumn{1}{p{4.085em}}{Training set} & \multicolumn{1}{p{4.835em}}{Validation set} & \multicolumn{1}{p{3.25em}}{Test set} & \multicolumn{1}{p{3.585em}|}{External set} \\
    \midrule
    \multicolumn{1}{|l|}{\multirow{3}[2]{*}{AVE bias minimizing}} & ECFP  & 35    & 9     & 32    & 0.996 & 0.689 & 0.714 & 0.735 \\
          & MOE   & 46    & 5     & 32    & 0.989 & 0.829 & 0.823 & 0.887 \\
          & Mordred & 500   & 10    & 64    & 1.000 & 0.788 & 0.786 & 0.874 \\
    \midrule
    \multirow{3}[2]{*}{Random} & ECFP  & 107   & 18    & 32    & 1.000 & 0.893 & 0.841 & 0.798 \\
          & MOE   & 248   & 9     & 32    & 1.000 & 0.913 & \cellcolor[rgb]{ 1,  1,  0}0.908 & 0.887 \\
          & Mordred & 141   & 71    & 32    & 1.000 & 0.917 & 0.879 & 0.862 \\
    \midrule
    \multirow{3}[2]{*}{Scaffold} & ECFP  & 327   & 13    & 32    & 1.000 & 0.858 & 0.867 & 0.759 \\
          & MOE   & 61    & 9     & 32    & 1.000 & 0.906 & \cellcolor[rgb]{ 1,  .753,  0}0.936 & 0.876 \\
          & Mordred & 572   & 18    & 32    & 1.000 & 0.854 & 0.894 & 0.872 \\
    \bottomrule
    \end{tabular}%
  \label{tab:rf_classif_perf}%
}
\end{table*}%

The best performing model overall was a random forest with MOE descriptor features, trained with a scaffold splitter; it contained 61 trees with maximum depth 9. It achieved a test set ROC AUC of 0.936, 87\% accuracy, precision 0.84, recall 0.89, negative predictive value 0.89 and Matthews correlation coefficient 0.74. These metrics significantly exceed those reported for the best previously published model\cite{montanari_flagging_2016}.

\subsection{Reproducibility of model performance results}
Models trained with the same parameters can yield different predictions and consequently different performance metrics, due to stochastic elements of the data partitioning, featurization and model training process. We assessed the variation in test set ROC AUC metrics across sets of 6 to 23 identically trained models for each combination of model type, splitter and feature type, using the parameters for the best performing models listed in tables \ref{tab:nn_classif_perf} and \ref{tab:rf_classif_perf} above. The results are shown in Figure \ref{fig:classif_perf_var} and in Table \ref{tab:classif_perf_var}. Generally, the random forest performance metrics were stable when the models were retrained, except when the AVE bias minimizing splitter was used. Neural networks showed greater variations in performance, due to the random dropouts used both at training and at prediction time.

\begin{table*}[tbp]
  \centering
  \caption{Ranges of test set ROC AUC scores across classification models retrained with best parameters for each model type, feature type, and data set splitting method.}
  \resizebox{\textwidth}{!}{%
    \begin{tabular}{|c|clc|cc|}
    \toprule
    \multicolumn{1}{|c|}{\multirow{2}[2]{*}{\textbf{Model type}}} & \multicolumn{1}{c}{\multirow{2}[2]{*}{\textbf{Splitter}}} & \multicolumn{1}{c}{\multirow{2}[2]{*}{\textbf{Features}}} & \multicolumn{1}{c|}{\multirow{2}[2]{*}{\textbf{Original test set ROC AUC}}} & \multicolumn{2}{c|}{\textbf{Test set ROC AUCs for retrained models}} \\
\cmidrule{5-6}          &       &       &       & \textbf{Mean} & \textbf{Standard deviation} \\
    \midrule
    \multicolumn{1}{|c|}{\multirow{9}[6]{*}{Neural \r network}} & \multicolumn{1}{c}{\multirow{3}[2]{*}{AVE bias minimizing}} & ECFP  & 0.770 & 0.767 & 0.022 \\
          &       & MOE   & 0.799 & 0.815 & 0.013 \\
          &       & Mordred  & 0.849 & 0.812 & 0.015 \\
\cmidrule{2-6}          & \multicolumn{1}{c}{\multirow{3}[2]{*}{Random}} & ECFP  & 0.846 & 0.846 & 0.007 \\
          &       & MOE   & 0.882 & 0.880 & 0.012 \\
          &       & Mordred  & 0.868 & 0.857 & 0.027 \\
\cmidrule{2-6}          & \multicolumn{1}{c}{\multirow{3}[2]{*}{Scaffold}} & ECFP  & 0.833 & 0.813 & 0.014 \\
          &       & MOE   & 0.907 & 0.893 & 0.010 \\
          &       & Mordred  & 0.851 & 0.822 & 0.027 \\
    \midrule
    \multicolumn{1}{|c|}{\multirow{9}[6]{*}{Random \r forest}} & \multicolumn{1}{c}{\multirow{3}[2]{*}{AVE bias minimizing}} & ECFP  & 0.823 & 0.742 & 0.026 \\
          &       & MOE   & 0.816 & 0.836 & 0.017 \\
          &       & Mordred  & 0.791 & 0.796 & 0.027 \\
\cmidrule{2-6}          & \multicolumn{1}{c}{\multirow{3}[2]{*}{Random}} & ECFP  & 0.849 & 0.858 & 0.004 \\
          &       & MOE   & 0.900 & 0.906 & 0.002 \\
          &       & Mordred  & 0.879 & 0.881 & 0.001 \\
\cmidrule{2-6}          & \multicolumn{1}{c}{\multirow{3}[2]{*}{Scaffold}} & ECFP  & 0.882 & 0.846 & 0.013 \\
          &       & MOE   & 0.925 & 0.906 & 0.007 \\
          &       & Mordred  & 0.891 & 0.868 & 0.009 \\
    \bottomrule
    \end{tabular}%
  \label{tab:classif_perf_var}%
}
\end{table*}%

\subsection{Comparison of results with single- and dual-threshold classification schemes}
The best previously reported model of BSEP inhibition\cite{montanari_flagging_2016} used a dual-threshold classification scheme that excluded weak BSEP inhibitors from the training and test sets. We performed an experiment to quantify the effect of such a scheme on reported performance. Following the same procedure as Montanari et al., we constructed a dual-threshold training and testing data set from which compounds with IC50s between 10 and 50 $\mu$M were excluded; compounds were classified as inhibitors if their IC50 was $< 10$ and as non-inhibitors if IC50 was $> 50 \mu$M. We used this dataset to train and test 220 neural network models, 113 with MOE and 107 with Mordred descriptor features, with a variety of hidden layer architectures. We trained and tested the same number of models, with equivalent feature sets and layer architectures, against our standard single-threshold classification dataset. For each split method, feature type and model architecture, we computed the change in test set ROC AUC scores between the models trained on dual- and single-threshold datasets.

\begin{figure*}[tp]
    \centering
    \includegraphics[width=.85\textwidth]{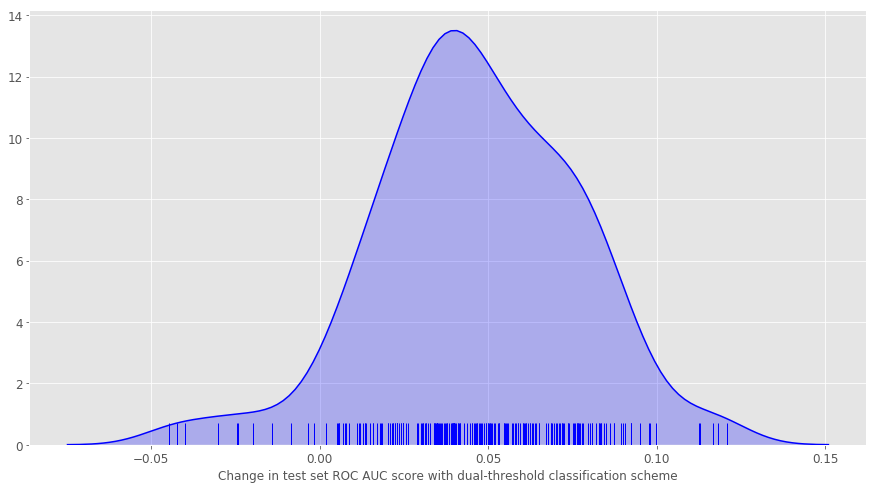}
    \caption{Distribution of increases in measured test set performance for neural network models trained and evaluated with dual-threshold classification datasets, compared to models of identical design trained and tested with single-threshold classification datasets.}
    \label{fig:dual_vs_single}
\end{figure*}

Figure \ref{fig:dual_vs_single} shows the frequency distribution of these differences. We see that using a dual-threshold dataset increases the measured ROC AUC in almost all (95\%) of cases; in the most extreme case, the AUC increased from 0.84 to 0.96. Table \ref{tab:dual_vs_single} shows the ROC AUC metrics for the models that performed best on the dual-threshold dataset for each split method and feature type, and corresponding models trained and tested on the single-threshold dataset. Models built using random splits tended to have greater increases in assessed performance than those trained with scaffold splits, averaging 0.06 for random vs 0.03 for scaffold splits. We find therefore that dual-threshold classification schemes lead to models with inflated performance estimates. Since compounds with intermediate IC50s for BSEP inhibition will be frequently encountered in drug discovery projects, such models can be expected to perform poorly in realistic applications.

\begin{table*}[htbp]
  \centering
  \caption{Comparison of best performing models trained and tested with dual- vs single-threshold datasets, for each splitting method and feature type.}
  \resizebox{\textwidth}{!}{%
    \begin{tabular}{|c|l|c|c|cc|}
    \toprule
    \multicolumn{1}{|c}{\multirow{2}[2]{*}{\textbf{Splitting method}}} & \multicolumn{1}{c}{\multirow{2}[2]{*}{\textbf{Feature type}}} & \multicolumn{1}{c}{\multirow{2}[2]{*}{\textbf{Hidden layers}}} & \multicolumn{1}{c}{\multirow{2}[2]{*}{\textbf{Layer sizes}}} & \multicolumn{2}{c|}{\textbf{Test set ROC AUC score}} \\
\cmidrule{5-6}          &       &       &       & \multicolumn{1}{p{6.165em}}{\textbf{Single threshold}} & \multicolumn{1}{p{4.415em}|}{\textbf{Dual threshold}} \\
    \midrule
    \multicolumn{1}{|c}{\multirow{2}[2]{*}{Random}} & MOE   & 1     & 8     & 0.871 & 0.936 \\
          & Mordred & 3     & 64,2,1 & 0.806 & 0.847 \\
    \midrule
    \multirow{2}[2]{*}{Scaffold} & MOE   & 3     & 16,4,2 & 0.898 & 0.920 \\
          & Mordred & 2     & 8,4   & 0.889 & 0.907 \\
    \bottomrule
    \end{tabular}%
  \label{tab:dual_vs_single}%
}
\end{table*}%

\subsection{Evaluation against external test set}
\begin{figure*}[tp]
    \centering
    \includegraphics[width=.85\textwidth]{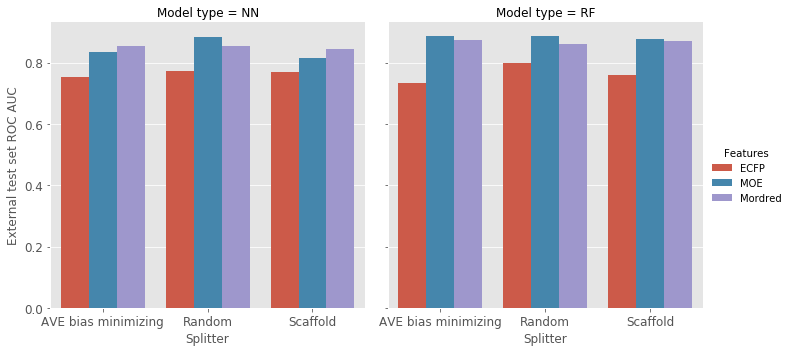}
    \caption{Performance of best classification models on external test set for each model type, feature type, and data set splitting method, computed as ROC AUC for external test set predictions. Model types are random forests (RF) and neural networks (NN).}
    \label{fig:classif_ext_perf}
\end{figure*}

\begin{figure*}[tp]
    \centering
    \includegraphics[width=.7\textwidth]{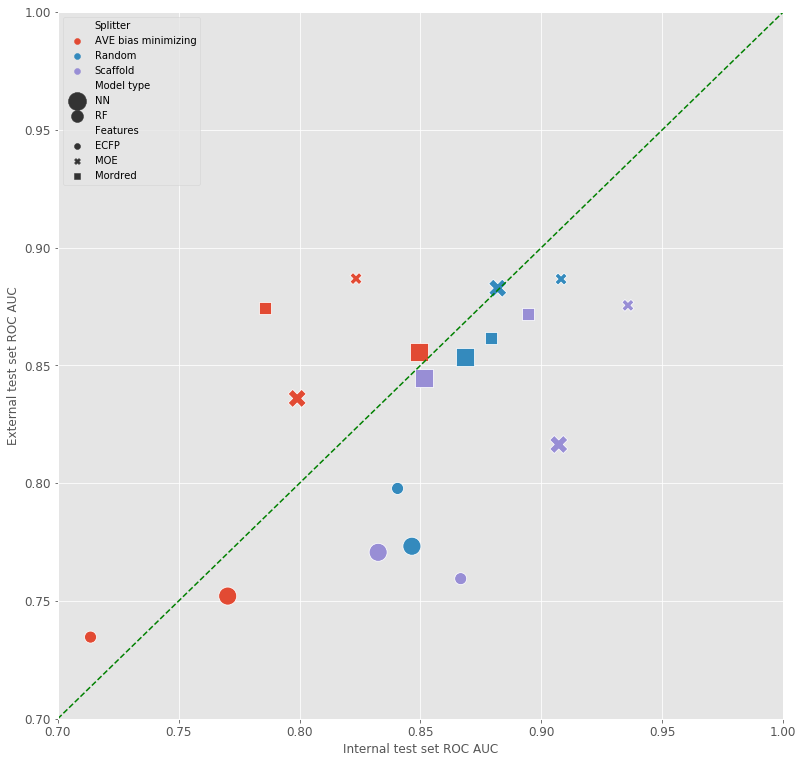}
    \caption{Comparison of classification model performance on external vs internal test sets. Markers are color coded by split type, shape coded by feature type, and sized by model type. The dashed line represents identical performance on the two datasets.}
    \label{fig:ext_vs_int}
\end{figure*}

For each combination of model type (neural network or random forest), splitting method and feature type, we selected our best performing classification model and used it to predict BSEP inhibition for a set of 366 compounds whose IC50s were measured by Warner et al.\cite{warner_mitigating_2012}. While the criterion used by Warner et al. to label a compound as an inhibitor was that its IC50 be less than 300 $\mu$M, we used a 50 $\mu$M threshold instead to maximize consistency between their labels and ours, as described in Methods. We computed the same performance metrics for the predictions as for our internal test set. The ROC AUC metrics for the external test set predictions are plotted in Figure \ref{fig:classif_ext_perf} and tabulated in Tables \ref{tab:nn_classif_perf} and \ref{tab:rf_classif_perf}. In Figure \ref{fig:ext_vs_int}, we compare the performance metrics for the predictions on the internal and external test sets.

Models based on MOE and Mordred descriptors generally performed well on the external dataset, in some cases with ROC AUCs exceeding those obtained on the internal test set. Since the external test compounds were selected to be chemically dissimilar from all compounds in our internal training and testing set, our results suggest that our classification models generalize well to new chemistry.

The type of split used to train a model did not greatly affect the absolute performance on the external dataset. However, it was notable that almost all of the best models trained with the AVE debiasing splitter performed better on the external dataset than on the internal test set. This was not true for models trained with the other splitters. The model with best overall performance on the external set was a MOE descriptor based random forest trained with an AVE debiasing split. This result provides some support for the hypothesis that removing AVE bias when partitioning datasets favors models that generalize beyond the training data.

\subsection{Performance of open-source BSEP classification models}
To provide classification models for use by the community, we trained a series of neural network models on a class-balanced dataset based on publicly available data for 660 compounds, with open-source Mordred descriptors as features. The input dataset was split into training, validation and test sets in the same way as the public-proprietary combined dataset, with 70\%, 15\% and 15\% of the compounds respectively, using either a random or a scaffold-based splitter. For each splitter, 124 models with different network architectures were trained. The best performing model was selected for each splitter based on the validation set ROC AUC score. The best models were then evaluated against the internal test set, and also on an external test set containing all GSK compounds from the public-proprietary combined dataset, filtered to exclude compounds structurally similar to any training, validation or test set compound.

Table \ref{tab:public_model_perf} shows the ROC AUC scores for the best models trained with each splitter type; additional performance metrics are shown in Supplemental Table S3. Generally speaking, the models trained with a random split produced higher ROC AUC scores than the scaffold split models when evaluated against the internal held-out test set. However, the best scaffold split model scored slightly better than the random split when assessed on the external test set. The relative generalization powers of the random and scaffold split models, as measured by performance on the external test set, depended on the choice of performance metric. While the two models had similar values for the ROC AUC, PRC AUC, accuracy and MCC metrics, the scaffold split model had notably better precision (0.83 vs 0.78), worse recall (0.75 vs 0.88) and worse NPV (0.62 vs 0.73) compared to the random split model.

\begin{table*}[htbp]
  \centering
  \caption{Performance (area under ROC curve) of best classification models trained against publicly available data.}
  \resizebox{\textwidth}{!}{%
    \begin{tabular}{|l|lccc|cccc|}
    \toprule
    \multicolumn{1}{|c|}{\multirow{2}[4]{*}{\textbf{Splitter}}} & \multicolumn{1}{c}{\multirow{2}[4]{*}{\textbf{Features}}} & \multicolumn{1}{c}{\multirow{2}[4]{*}{\textbf{\# Hidden layers}}} & \multicolumn{1}{c}{\multirow{2}[4]{*}{\textbf{Hidden layer sizes}}} & \multicolumn{1}{c|}{\multirow{2}[4]{*}{\textbf{Training epochs}}} & \multicolumn{4}{c|}{\textbf{ROC AUC}} \\
\cmidrule{6-9}          &       &       &       &       & \multicolumn{1}{p{5em}}{Training set} & \multicolumn{1}{p{5em}}{Validation set} & \multicolumn{1}{p{5em}}{Test set} & \multicolumn{1}{p{5em}|}{External set} \\
    \midrule
    Random & Mordred & 3     & 16,9,5 & 264   & 0.995 & 0.901 & 0.912 & 0.802 \\
    \midrule
    Scaffold & Mordred & 3     & 16,12,3 & 359   & 1     & 0.875 & 0.742 & 0.812 \\
    \bottomrule
    \end{tabular}%
  }
  \label{tab:public_model_perf}%
\end{table*}%

Comparing performance of models trained with public vs proprietary data is problematic, given the lack of a common test set and the limited range of model and feature types tested with public data. Nevertheless, we observe that, by most metrics, our public-data models perform almost as well as our proprietary-data models against their respective external test sets. We also noted that our models performed better than the best previously published model on their internal test sets. Our best public-data models for each split and the code and data used to generate them are available as supplementary data, for use with our open source AMPL modeling pipeline.

\subsection{Performance of BSEP regression (pIC50) models}
\begin{figure*}[tp]
    \centering
    \includegraphics[width=.85\textwidth]{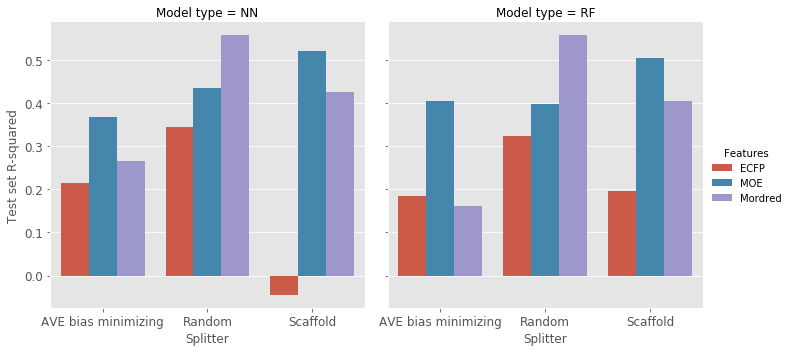}
    \caption{Performance of best regression models for each model type, feature type, and data set splitting method, measured as $R^2$ for test set predictions.}
    \label{fig:regr_perf}
\end{figure*}

Figure \ref{fig:regr_perf} summarizes the performance, measured by test set $R^2$ scores, of the best regression models for each combination 
of model type, split method, and feature type. Characteristics of the models and $R^2$ scores for each subset are shown 
in Tables \ref{tab:nn_regr_perf} and \ref{tab:rf_regr_perf}, while additional performance metrics are documented in Supplemental Tables S4 and S5. As with classification models, the AVE debiasing splitter resulted in lower reported performance metrics than random or scaffold splitting, and models featurized using chemical descriptors usually performed much better than models based on ECFP4 fingerprints.  In fact, the “best” neural network model (in terms of validation set $R^2$) trained with a scaffold split and ECFP4 fingerprints yielded a negative $R^2$ value on the test set, indicating that the residual variance of its predictions was greater than the sample variance of the data.

In 4 out of 6 cases, models using MOE descriptors outperformed Mordred-based models of the same type trained with the same splitter; however, the models with the best overall reported metrics were based on Mordred descriptors, with the top neural network model slightly outperforming its random forest counterpart. It had a test set $R^2$ of 0.557, mean absolute error (MAE) 0.37, and RMS error 0.577. Since the predicted pIC50 values are log space measurements, the MAE corresponds to a $10^{0.37}$ or 2.3-fold error in predicted IC50s. This result is comparable to the observed variation between experimental replicates in our laboratory, suggesting that further increases in accuracy may be difficult to achieve. We note however that, since this model was trained using a random split, its performance metrics may be inflated due to similarities between some test and training compounds.

\begin{table*}[htbp]
  \centering
  \caption{Model hyperparameters and $R^2$ scores of best neural network regression models. Test set $R^2$ values for best performing models 
within each splitter class are highlighted in yellow; the best overall $R^2$ is highlighted in orange.}
  \resizebox{\textwidth}{!}{%
    \begin{tabular}{|llccc|ccc|}
    \toprule
    \multicolumn{1}{|c}{\multirow{2}[4]{*}{\textbf{Splitter}}} & \multicolumn{1}{c}{\multirow{2}[4]{*}{\textbf{Features}}} & \multicolumn{1}{c}{\multirow{2}[4]{*}{\textbf{\# hidden layers}}} & \multicolumn{1}{c}{\multirow{2}[4]{*}{\textbf{Hidden layer sizes}}} & \multicolumn{1}{c|}{\multirow{2}[4]{*}{\textbf{Training epochs}}} & \multicolumn{3}{c|}{\textbf{$R^2$}} \\
\cmidrule{6-8}          &       &       &       &       & \multicolumn{1}{p{5em}}{Training set} & \multicolumn{1}{p{5em}}{Validation set} & \multicolumn{1}{p{5em}|}{Test set} \\
    \midrule
    \multicolumn{1}{|l}{\multirow{3}[2]{*}{AVE bias minimizing}} & ECFP  & 1     & 16    & 16    & 0.620 & 0.209 & 0.214 \\
          & MOE   & 3     & 128,32,16 & 125   & 0.841 & 0.498 & 0.368 \\
          & Mordred & 2     & 64,4  & 202   & 0.813 & 0.503 & 0.264 \\
    \midrule
    \multirow{3}[2]{*}{Random} & ECFP  & 1     & 16    & 14    & 0.577 & 0.351 & 0.345 \\
          & MOE   & 2     & 32,15 & 349   & 0.891 & 0.583 & 0.435 \\
          & Mordred & 2     & 64,12 & 478   & 0.927 & 0.601 & \cellcolor[rgb]{ 1,  .753,  0}0.557 \\
    \midrule
    \multirow{3}[2]{*}{Scaffold} & ECFP  & 3     & 16,15,13 & 229   & 0.814 & 0.360 & -0.045 \\
          & MOE   & 1     & 128   & 133   & 0.673 & 0.554 & \cellcolor[rgb]{ 1,  1,  0}0.521 \\
          & Mordred & 2     & 64,14 & 24    & 0.608 & 0.506 & 0.426 \\
    \bottomrule
    \end{tabular}%
  }
  \label{tab:nn_regr_perf}%
\end{table*}%

\begin{table*}[htbp]
  \centering
  \caption{Model hyperparameters and $R^2$ scores of best random forest regression models.}
  \resizebox{\textwidth}{!}{%
    \begin{tabular}{|llccc|ccc|}
    \toprule
    \multicolumn{1}{|c}{\multirow{2}[4]{*}{\textbf{Splitter}}} & \multicolumn{1}{c}{\multirow{2}[4]{*}{\textbf{Features}}} & \multirow{2}[4]{*}{\textbf{\# Trees}} & \multirow{2}[4]{*}{\textbf{Max depth}} & \multicolumn{1}{c|}{\multirow{2}[4]{*}{\textbf{Max features per split}}} & \multicolumn{3}{c|}{\textbf{$R^2$}} \\
\cmidrule{6-8}          &       &       &       &       & \multicolumn{1}{p{5em}}{Training set} & \multicolumn{1}{p{5em}}{Validation set} & \multicolumn{1}{p{5em}|}{Test set} \\
    \midrule
    \multicolumn{1}{|l}{\multirow{3}[2]{*}{AVE bias minimizing}} & ECFP  & 81    & 18    & 32    & 0.871 & 0.233 & 0.184 \\
          & MOE   & 61    & 18    & 32    & 0.945 & 0.441 & \cellcolor[rgb]{ 1,  1,  0}0.405 \\
          & Mordred & 11    & 26    & 32    & 0.903 & 0.396 & 0.160 \\
    \midrule
    \multirow{3}[2]{*}{Random} & ECFP  & 46    & 36    & 32    & 0.906 & 0.392 & 0.324 \\
          & MOE   & 61    & 100   & 32    & 0.943 & 0.618 & 0.397 \\
          & Mordred & 107   & 13    & 32    & 0.922 & 0.559 & 0.557 \\
    \midrule
    \multirow{3}[2]{*}{Scaffold} & ECFP  & 20    & 100   & 32    & 0.890 & 0.339 & 0.195 \\
          & MOE   & 46    & 13    & 32    & 0.934 & 0.529 & 0.504 \\
          & Mordred & 26    & 13    & 32    & 0.918 & 0.459 & 0.405 \\
    \bottomrule
    \end{tabular}%
  }
  \label{tab:rf_regr_perf}%
\end{table*}%

\begin{figure*}[tp]
    \centering
    \includegraphics[width=.75\textwidth]{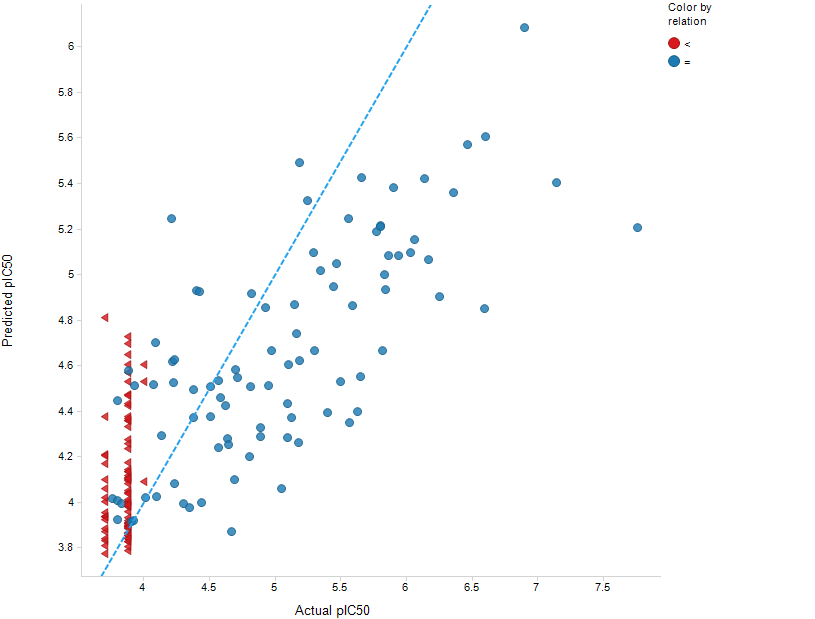}
    \caption{Predicted vs actual test set pIC50 values for the best overall regression model. Compounds whose actual pIC50 values were left-censored are represented by triangles. The dashed line represents the identity relation.}
    \label{fig:pred_vs_actual}
\end{figure*}

In general, regression models for dose-response assays are challenging to fit, due to the limited range of compound concentrations tested. 
Compounds that don’t produce at least 50\% inhibition at the maximum concentration $C_{max}$ are reported as having IC50 $>C_{max}$; 
the corresponding pIC50 values are thus left-censored at $-\log_{10} C_{max}$. It is also possible for dose-response datasets to be 
right-censored; for example, Warner et al. tested compounds at a minimum concentration of 10 $\mu$M, so the maximum pIC50 reported in 
their dataset is 5. Additional complications result when, as in our combined dataset, the data contain a mixture of values censored at 
different thresholds. The situation is illustrated in Figure \ref{fig:pred_vs_actual}, which shows the predicted pIC50 
values for our best performing model plotted against the actual measured values for the test dataset compounds. In our combined dataset, the measured IC50 values were typically censored at either 133 $\mu$M (from the Morgan et al. dataset) or 200 $\mu$M (from the GSK subset).  The two vertical strings of points at the lower left corner of the plot represent the spread of predictions for the compounds with censored measurements. 

Typically, researchers follow one of two approaches to fit regression models to pIC50 data: either omit the censored data points from the training and testing sets, or add an arbitrary negative offset to the censored pIC50 values. For the sake of interpretability, we did not use the latter approach here; and in our experience, removing censored data points from the training set yields less predictive regression models. A more robust approach to quantitatively predict BSEP inhibition would use a hybrid model, combining a classifier to predict whether the pIC50 was above or below the censoring threshold, together with a regression model trained only on uncensored data that would be applied only to compounds predicted to have pIC50s above the threshold. Alternatively, one could fit a Tobit model\cite{amemiya_tobit_1984}, which predicts the mean and variance of a Gaussian distribution underlying the observed censored pIC50 values for each compound. Both of these approaches are being actively investigated by our group.

%% file: sections/Conclusion.tex
In this work, we have presented our results from using an automated data-driven modeling pipeline (AMPL) to train and evaluate over 15,500 classification and regression models of BSEP inhibition, based on a combined proprietary and public dataset containing IC50 data for 1,149 compounds. By using the hyperparameter search capability of AMPL, we were able to test a wide variety of combinations of model types, dataset splitting strategies, chemical featurization methods, and model parameters. When evaluated against our internal held-out test dataset, our best model significantly outperformed the best previously reported model across all standard metrics for classification models. Its superior performance is especially remarkable, given our demonstration that the dual-threshold classification scheme used to train and test the competing model leads to inflated performance estimates. While our best model and the one reported by Montanari et al. both used a random forest algorithm with MOE descriptors as features, it appears that the random forest implementation used by AMPL (based on the scikit-learn Python package) allows variation of a wider range of model parameters than the Weka software used by Montanari et al. In addition, the hyperparameter search tools in AMPL allowed us to test many combinations of parameters to find an optimal set.

For each combination of model type, splitter and feature type, we tested the best performing classification model against a publicly available external dataset to assess its power to generalize to novel chemistry. We found that models based on MOE descriptors generally performed best when evaluated against both our internal test set and the external dataset; however, models using Mordred descriptors performed almost as well, and outperformed MOE descriptors in a few cases.

Our study found much variation in assessed performance between models trained using different dataset splitting strategies, with the AVE debiasing splitter providing more stringent tests of model performance on our internal test set. When we compared performance on the external dataset to the internal test set, most models trained with the AVE splitter scored better on the external set, despite the fact that the external set was filtered to exclude compounds with structural similarity to the training data. While this result suggests that AVE debiasing may help produce models that generalize better, we note that the model that scored best on the external set (which was trained with an AVE split) only slightly outperformed another model that used a random split. We plan in future work to further investigate the factors affecting generalization performance, using a wider variety of datasets.

To our knowledge, our study is the first to report results from fitting regression models to predict pIC50 values for BSEP inhibition. We found that our best model achieved a reasonable coefficient of determination ($R^2$), the most commonly used metric for regression models. More importantly for practical purposes, its mean absolute error (MAE) corresponds to a 2.3-fold variation in predicted IC50 values. This is on the same scale as the observed variation across experimental replicate IC50 measurements. We are excited by this result, as it means that the model can be used to predict input values for simulations of drug-induced liver injury (e.g., DILI-sym), as well as in computational pipelines for in silico drug discovery. We look forward to the availability of large datasets measuring inhibition of other bile salt transporters, such as NTCP, MRP3 and MRP4; this will make possible the construction of accurate models for these additional factors which are important predictors of cholestatic drug-induced liver injury.

In order to make models of BSEP inhibition widely available and encourage development of improved models, we trained a series of models on publicly available data and evaluated their generalization power on a proprietary compound set. The best performing models, along with code and data used to generate them, are released as supplementary data with this article, to be used with the open-source AMPL pipeline software. We hope that other researchers will use the AMPL platform to develop even better models, by training a wider variety of model architectures against larger, more diverse datasets, and contribute their models to the community.

%% file: sections/Supplement.tex
\subsection{Models, Software and Training Data}
Models trained with publicly available data, along with the data and code used to train and test them, are available at the AMPL GitHub repository: 
\url{https://github.com/ATOMconsortium/AMPL}, in subdirectory \texttt{atomsci/ddm/examples/BSEP}. Instructions for running the models are in the file 
\texttt{README.md} in that directory.

\subsection{Maximum Likelihood Mean Estimation for Partially Censored Data}
Estimates of IC50 drug concentrations for a particular biological activity, such as BSEP inhibition, are restricted by the range of 
concentrations over which activities are measured, [$C_{min}, C_{max}$]. If the target activity at $C_{max}$ is not reduced by at 
least 50\%, the IC50 is typically reported as $>C_{max}$ and the corresponding pIC50 as $<-\log_{10}C_{max}$; that is, 
the pIC50 observations are left-censored. Likewise, observations may be right-censored if $>50\%$ inhibition occurs at all concentrations tested.

In our BSEP inhibition training and testing dataset, we had replicate pIC50 measurements reported as $x_i, i=1 \ldots n$, for most compounds; 
these were combined to produce a single pIC50 value for each compound. In many cases, some or all of the replicate measurements 
were censored. We treated the replicates as samples from an underlying Gaussian distribution with mean $\mu$ and standard deviation $\sigma$. 
We estimated $\mu$ as follows for the four possible cases:
\begin{itemize}
\item All replicates left-censored: 
\[
\hat{\mu} = \min_{i = 1 \ldots n} x_i
\]
\item All replicates right-censored:
\[
\hat{\mu} = \max_{i = 1 \ldots n} x_i 
\]
\item All replicates uncensored: 
\[
\hat{\mu} = \frac{1}{n} \sum_{i = 1}^{n} x_i
\]
\item Some replicates censored, some uncensored: 
\[
\hat{\mu} = \argmax_{\mu} \loglik(\mu,\sigma;x)
\]
where $\loglik(\mu,\sigma;x)$ is the log likelihood function defined as follows:
\end{itemize}
\begin{align*}
\loglik(\mu,\sigma ; x) &= \sum_{i \in U} \log \phi \left(\frac{x_i - \mu}{\sigma} \right) \\
                      &+ \sum_{i \in L} \log \Phi \left(\frac{x_i - \mu}{\sigma} \right) \\
                      &+ \sum_{i \in R} \log \left[ 1 - \Phi \left(\frac{x_i - \mu}{\sigma}\right) \right]
\end{align*}

In the above, $U, L$ and $R$ are respectively the sets of replicate indices for uncensored, left- and right-censored measurements; 
$\phi$ is the probability density function for the standard Gaussian distribution; 
and $\Phi$ is the cumulative distribution function for the standard Gaussian distribution. We estimated a common standard 
deviation parameter $\sigma$ for all compounds by computing the RMS deviation of all uncensored replicate measurements from their 
compound-specific means, across compounds with at least two uncensored replicates. The maximum likelihood estimate was obtained using 
Brent’s algorithm, as implemented in package \texttt{scipy.optimize} function \texttt{minimize\_scalar}.

\subsection{Modifications to AVE Debiasing Algorithm}
To partition training, validation and test datasets, we implemented a modified version of the asymmetric validation embedding (AVE) bias 
minimization algorithm presented by Wallach et al.\cite{wallach_most_2018}. The original algorithm seeks to minimize the AVE bias, 
defined as follows: 
Let $T$ and $V$ be proposed training 
and validation sets of compounds, and let $\ta, \ti, \va$ and $\vi$ be the sets of “active” and “inactive” 
compounds within the training and validation sets. For our model, actives correspond to BSEP inhibitors and inactives 
to non-inhibitors. Define the “nearest neighbor function”:
\begin{equation*}
S(V,T,d) = \frac{1}{|V|} \sum_{v \in V} I \left[ D_{nn} (v,T) < d \right]
\end{equation*}
where $D_{nn} (v,T)$ is the Tanimoto distance of compound $v$ from its nearest neighbor in $T$, and $I$ is the indicator function.
Note that, when the sets $T$ and $V$ are fixed, $S(V,T,d)$ is the cumulative empirical distribution function of $D_{nn}(v,T)$.
Next, fix a set of distance thresholds $D$; Wallach et al. use $D = \{0, 0.01, \ldots ,1\}$ for Tanimoto distance. 
Then, define the average CDF over this set of distances:
\begin{equation*}
H(V,T) = \frac{1}{|D|} \sum_{d \in D} S(V,T,d)
\end{equation*}
Finally, the AVE bias is defined as:
\begin{align*}
B(\va,\vi,\ta,\ti) &= [H(\va,\ta) - H(\va,\ti)] \\
 &+ [H(\vi,\ti) - H(\vi,\ta)]
\end{align*}

The first bracketed term is similar to the MUV bias defined by Rohrer et al.\cite{rohrer_maximum_2009}; 
it represents the tendency for validation set active 
compounds to cluster with training set actives. The second bracketed term measures the tendency for validation set inactives to cluster 
with training set inactives. Either of these terms can be negative; therefore a training/validation split with a negative contribution from the 
inactive term and positive contribution from the active term can appear to have low bias. We viewed this as a defect in the AVE debiasing 
algorithm, since it favors classification algorithms that memorize the active compounds in the training data. 
In our implementation, we used a modified version of the AVE bias that enforces positive contributions for both the active and inactive compounds:
\begin{align*}
B'(\va,\vi,\ta,\ti) &= |H(\va,\ta) - H(\va,\ti)| \\
  &+ |H(\vi,\ti) - H(\vi,\ta)|
\end{align*}

Our second modification to the AVE debiasing algorithm was to support chemical descriptor features as an alternative to the fingerprint bit 
vectors supported by the original algorithm. For models using descriptors as features, we use Euclidean distance between feature vectors 
to compute nearest-neighbor distances. Unlike Tanimoto distances, these distances are not bounded in the range [0,1], so an alternative 
set of distances $D$ must be used to compute the average CDF $H(V,T)$. 
We chose a similar set to the one used by Rohrer et al. to compute the MUV bias: a set of 100 evenly spaced distances 
from 0 to $3 \cdot \mathrm{median}(D_{nn})$, where the median was computed over the distances from all compounds to 
their nearest neighbors in the unpartitioned dataset.

In most other respects, our version of the AVE debiasing algorithm follows the implementation provided by Wallach et al.; it generates a set of random 2-way splits, then uses a genetic algorithm to combine and mutate the splits iteratively until the computed AVE bias falls below some threshold. To generate the 3-way training/validation/test splits we use for model building and evaluation, we perform an initial 2-way split to select the held-out test set, followed by a second split on the remaining compounds to separate the training and validation sets. The final training and test sets may have a small residual AVE bias due to the removal of the validation set compounds, but in our experience it is not large enough to be of concern.

%% file: sections/SuppTables.tex
\thispagestyle{empty}
\begin{landscape}
\subsection{Supplemental Tables}
  \centering
  \resizebox{\linewidth}{!}{%
    \begin{tabular}{|l|lccc|cccc|cccc|cccc|cccc|cccc|cccc|cccr|}
    \toprule
    \multicolumn{1}{|c|}{\multirow{2}[4]{*}{\textbf{Splitter}}} & \multicolumn{1}{c}{\multirow{2}[4]{*}{\textbf{Features}}} & \multicolumn{1}{c}{\multirow{2}[4]{*}{\textbf{\# Hidden layers}}} & \multicolumn{1}{c}{\multirow{2}[4]{*}{\textbf{Hidden layer sizes}}} & \multicolumn{1}{c|}{\multirow{2}[4]{*}{\textbf{Training epochs}}} & \multicolumn{4}{c|}{\textbf{ROC AUC}} & \multicolumn{4}{c|}{\textbf{PRC AUC}} & \multicolumn{4}{c|}{\textbf{Accuracy}} & \multicolumn{4}{c|}{\textbf{Precision}} & \multicolumn{4}{c|}{\textbf{Recall}} & \multicolumn{4}{c|}{\textbf{NPV}} & \multicolumn{4}{c|}{\textbf{MCC}} \\
\cmidrule{6-33}          &       &       &       &       & \multicolumn{1}{p{4.085em}}{Training set} & \multicolumn{1}{p{4.835em}}{Validation set} & \multicolumn{1}{p{3.25em}}{Test set} & \multicolumn{1}{p{3.585em}|}{External set} & \multicolumn{1}{p{5em}}{Training set} & \multicolumn{1}{p{5em}}{Validation set} & \multicolumn{1}{p{5em}}{Test set} & \multicolumn{1}{p{5em}|}{External set} & \multicolumn{1}{p{5em}}{Training set} & \multicolumn{1}{p{5em}}{Validation set} & \multicolumn{1}{p{5em}}{Test set} & \multicolumn{1}{p{5em}|}{External set} & \multicolumn{1}{p{5em}}{Training set} & \multicolumn{1}{p{5em}}{Validation set} & \multicolumn{1}{p{5em}}{Test set} & \multicolumn{1}{p{5em}|}{External set} & \multicolumn{1}{p{5em}}{Training set} & \multicolumn{1}{p{5em}}{Validation set} & \multicolumn{1}{p{5em}}{Test set} & \multicolumn{1}{p{5em}|}{External set} & \multicolumn{1}{p{5em}}{Training set} & \multicolumn{1}{p{5em}}{Validation set} & \multicolumn{1}{p{5em}}{Test set} & \multicolumn{1}{p{5em}|}{External set} & \multicolumn{1}{p{5em}}{Training set} & \multicolumn{1}{p{5em}}{Validation set} & \multicolumn{1}{p{5em}}{Test set} & \multicolumn{1}{p{5em}|}{External set} \\
    \midrule
    \multicolumn{1}{|l|}{\multirow{3}[2]{*}{AVE bias minimizing}} & ECFP  & 2     & 64,8  & 2     & 0.952 & 0.699 & 0.770 & 0.752 & 0.945 & 0.659 & 0.716 & 0.627 & 0.866 & 0.661 & 0.717 & 0.678 & 0.929 & 0.705 & 0.745 & 0.867 & 0.752 & 0.419 & 0.528 & 0.101 & 0.831 & 0.645 & 0.704 & 0.670 & 0.733 & 0.310 & 0.418 & 0.223 \\
          & MOE   & 2     & 12,7  & 155   & 0.990 & 0.866 & 0.799 & 0.836 & 0.989 & 0.828 & 0.742 & 0.707 & 0.958 & 0.787 & 0.737 & 0.767 & 0.991 & 0.790 & \cellcolor[rgb]{ 1,  1,  0}0.784 & 0.734 & 0.914 & 0.690 & 0.548 & 0.535 & 0.936 & 0.784 & 0.716 & 0.779 & 0.917 & 0.562 & 0.464 & 0.469 \\
          & Mordred & 2     & 15,5  & 10    & 0.926 & 0.834 & \cellcolor[rgb]{ 1,  1,  0}0.849 & 0.856 & 0.900 & 0.782 & 0.765 & 0.721 & 0.847 & 0.755 & \cellcolor[rgb]{ 1,  1,  0}0.794 & 0.801 & 0.803 & 0.678 & 0.714 & 0.759 & 0.865 & 0.831 & \cellcolor[rgb]{ 1,  1,  0}0.878 & 0.636 & 0.887 & 0.842 & \cellcolor[rgb]{ 1,  1,  0}0.886 & 0.818 & 0.694 & 0.523 & \cellcolor[rgb]{ 1,  1,  0}0.604 & 0.551 \\
    \midrule
    \multirow{3}[2]{*}{Random} & ECFP  & 3     & 14,12,7 & 13    & 0.959 & 0.913 & 0.846 & 0.773 & 0.952 & 0.893 & 0.847 & 0.668 & 0.876 & 0.831 & 0.780 & 0.694 & 0.894 & 0.852 & 0.781 & 0.707 & 0.814 & 0.722 & 0.722 & 0.225 & 0.863 & 0.820 & 0.780 & 0.692 & 0.748 & 0.652 & 0.556 & 0.264 \\
          & MOE   & 3     & 15,6,3 & 74    & 0.967 & 0.911 & 0.882 & 0.883 & 0.959 & 0.853 & 0.833 & 0.756 & 0.925 & 0.860 & \cellcolor[rgb]{ 1,  1,  0}0.827 & 0.764 & 0.933 & 0.911 & \cellcolor[rgb]{ 1,  1,  0}0.838 & 0.765 & 0.897 & 0.729 & 0.750 & 0.481 & 0.919 & 0.836 & 0.819 & 0.764 & 0.849 & 0.712 & \cellcolor[rgb]{ 1,  1,  0}0.647 & 0.460 \\
          & Mordred & 3     & 13,5,3 & 96    & 0.969 & 0.950 & 0.868 & 0.854 & 0.958 & 0.932 & 0.868 & 0.770 & 0.949 & 0.849 & 0.780 & 0.798 & 0.970 & 0.883 & 0.824 & 0.802 & 0.912 & 0.736 & 0.683 & 0.566 & 0.935 & 0.830 & 0.752 & 0.796 & 0.897 & 0.689 & 0.563 & 0.542 \\
    \midrule
    \multirow{3}[2]{*}{Scaffold} & ECFP  & 2     & 16,11 & 4     & 0.918 & 0.883 & 0.833 & 0.771 & 0.870 & 0.906 & 0.870 & 0.659 & 0.770 & 0.593 & 0.584 & 0.680 & 0.932 & 0.966 & \cellcolor[rgb]{ 1,  .753,  0}1.000 & 1.000 & 0.441 & 0.289 & 0.242 & 0.093 & 0.733 & 0.517 & 0.520 & 0.669 & 0.529 & 0.365 & 0.355 & 0.250 \\
          & MOE   & 3     & 32,16,4 & 128   & 0.969 & 0.905 & 0.907 & 0.816 & 0.969 & 0.916 & 0.899 & 0.689 & 0.960 & 0.814 & 0.815 & 0.767 & 0.990 & 0.875 & 0.819 & 0.700 & 0.910 & 0.786 & 0.800 & 0.597 & 0.943 & 0.750 & 0.811 & 0.796 & 0.918 & 0.631 & 0.630 & 0.476 \\
          & Mordred & 3     & 64,16,1 & 17    & 0.914 & 0.899 & 0.851 & 0.845 & 0.803 & 0.913 & 0.827 & 0.692 & 0.866 & 0.791 & 0.740 & 0.790 & 0.815 & 0.918 & 0.798 & 0.728 & 0.847 & 0.691 & 0.705 & 0.643 & 0.900 & 0.697 & 0.685 & 0.817 & 0.720 & 0.613 & 0.485 & 0.529 \\
    \bottomrule
    \end{tabular}%
}
\captionof*{table}{Table S1. Full performance metrics for best neural network classification models.}
\label{tab:full_nn_classif_metrics}%

  \thispagestyle{empty}
  \centering
  \resizebox{\linewidth}{!}{%
    \begin{tabular}{|l|lccc|cccc|cccc|cccc|cccc|cccc|cccc|cccr|}
    \toprule
    \multicolumn{1}{|c|}{\multirow{2}[4]{*}{\textbf{Splitter}}} & \multicolumn{1}{c}{\multirow{2}[4]{*}{\textbf{Features}}} & \multirow{2}[4]{*}{\textbf{\# Trees}} & \multirow{2}[4]{*}{\textbf{Max depth}} & \multicolumn{1}{c|}{\multirow{2}[4]{*}{\textbf{Max features per split}}} & \multicolumn{4}{c|}{\textbf{ROC AUC}} & \multicolumn{4}{c|}{\textbf{PRC AUC}} & \multicolumn{4}{c|}{\textbf{Accuracy}} & \multicolumn{4}{c|}{\textbf{Precision}} & \multicolumn{4}{c|}{\textbf{Recall}} & \multicolumn{4}{c|}{\textbf{NPV}} & \multicolumn{4}{c|}{\textbf{MCC}} \\
\cmidrule{6-33}          &       &       &       &       & \multicolumn{1}{p{4.085em}}{Training set} & \multicolumn{1}{p{4.835em}}{Validation set} & \multicolumn{1}{p{3.25em}}{Test set} & \multicolumn{1}{p{3.585em}|}{External set} & \multicolumn{1}{p{5em}}{Training set} & \multicolumn{1}{p{5em}}{Validation set} & \multicolumn{1}{p{5em}}{Test set} & \multicolumn{1}{p{5em}|}{External set} & \multicolumn{1}{p{5em}}{Training set} & \multicolumn{1}{p{5em}}{Validation set} & \multicolumn{1}{p{5em}}{Test set} & \multicolumn{1}{p{5em}|}{External set} & \multicolumn{1}{p{5em}}{Training set} & \multicolumn{1}{p{5em}}{Validation set} & \multicolumn{1}{p{5em}}{Test set} & \multicolumn{1}{p{5em}|}{External set} & \multicolumn{1}{p{5em}}{Training set} & \multicolumn{1}{p{5em}}{Validation set} & \multicolumn{1}{p{5em}}{Test set} & \multicolumn{1}{p{5em}|}{External set} & \multicolumn{1}{p{5em}}{Training set} & \multicolumn{1}{p{5em}}{Validation set} & \multicolumn{1}{p{5em}}{Test set} & \multicolumn{1}{p{5em}|}{External set} & \multicolumn{1}{p{5em}}{Training set} & \multicolumn{1}{p{5em}}{Validation set} & \multicolumn{1}{p{5em}}{Test set} & \multicolumn{1}{p{5em}|}{External set} \\
    \midrule
    \multicolumn{1}{|l|}{\multirow{3}[2]{*}{AVE bias minimizing}} & ECFP  & 35    & 9     & 32    & 0.996 & 0.689 & 0.714 & 0.735 & 0.995 & 0.650 & 0.636 & 0.616 & 0.946 & 0.636 & 0.663 & 0.710 & 0.997 & 0.652 & 0.682 & 0.709 & 0.879 & 0.405 & 0.417 & 0.302 & 0.913 & 0.630 & 0.656 & 0.711 & 0.893 & 0.255 & 0.301 & 0.314 \\
          & MOE   & 46    & 5     & 32    & 0.989 & 0.829 & 0.823 & 0.887 & 0.986 & 0.783 & \cellcolor[rgb]{ 1,  1,  0}0.785 & 0.817 & 0.952 & 0.750 & 0.754 & 0.808 & 0.933 & 0.688 & 0.722 & 0.804 & 0.960 & 0.775 & 0.712 & 0.605 & 0.968 & 0.810 & 0.779 & 0.810 & 0.903 & 0.501 & 0.500 & 0.567 \\
          & Mordred & 500   & 10    & 64    & 1.000 & 0.788 & 0.786 & 0.874 & 1.000 & 0.746 & 0.756 & 0.780 & 0.999 & 0.706 & 0.700 & 0.781 & 0.997 & 0.642 & 0.662 & 0.753 & 1.000 & 0.732 & 0.635 & 0.566 & 1.000 & 0.768 & 0.727 & 0.792 & 0.997 & 0.414 & 0.387 & 0.503 \\
    \midrule
    \multirow{3}[2]{*}{Random} & ECFP  & 107   & 18    & 32    & 1.000 & 0.893 & 0.841 & 0.798 & 1.000 & 0.855 & 0.835 & 0.680 & 0.990 & 0.820 & 0.763 & 0.732 & 1.000 & 0.815 & 0.797 & 0.746 & 0.977 & 0.736 & 0.646 & 0.364 & 0.983 & 0.822 & 0.743 & 0.729 & 0.980 & 0.627 & 0.523 & 0.376 \\
          & MOE   & 248   & 9     & 32    & 1.000 & 0.913 & \cellcolor[rgb]{ 1,  1,  0}0.908 & 0.887 & 1.000 & 0.866 & \cellcolor[rgb]{ 1,  1,  0}0.896 & 0.810 & 0.994 & 0.831 & 0.815 & 0.795 & 0.986 & 0.825 & 0.806 & 0.770 & 1.000 & 0.743 & 0.763 & 0.597 & 1.000 & 0.835 & \cellcolor[rgb]{ 1,  1,  0}0.822 & 0.804 & 0.988 & 0.648 & 0.623 & 0.535 \\
          & Mordred & 141   & 71    & 32    & 1.000 & 0.917 & 0.879 & 0.862 & 1.000 & 0.904 & 0.869 & 0.763 & 0.998 & 0.855 & 0.815 & 0.776 & 0.997 & 0.841 & 0.821 & 0.764 & 0.997 & 0.806 & \cellcolor[rgb]{ 1,  1,  0}0.780 & 0.527 & 0.998 & 0.864 & 0.811 & 0.780 & 0.995 & 0.700 & 0.629 & 0.488 \\
    \midrule
    \multirow{3}[2]{*}{Scaffold} & ECFP  & 327   & 13    & 32    & 1.000 & 0.858 & 0.867 & 0.759 & 0.999 & 0.885 & 0.901 & 0.646 & 0.966 & 0.663 & 0.688 & 0.705 & 1.000 & 0.953 & 0.936 & 0.784 & 0.914 & 0.423 & 0.463 & 0.225 & 0.948 & 0.566 & 0.595 & 0.696 & 0.931 & 0.454 & 0.475 & 0.303 \\
          & MOE   & 61    & 9     & 32    & 1.000 & 0.906 & \cellcolor[rgb]{ 1,  .753,  0}0.936 & 0.876 & 1.000 & 0.926 & \cellcolor[rgb]{ 1,  .753,  0}0.935 & 0.786 & 0.990 & 0.820 & \cellcolor[rgb]{ 1,  .753,  0}0.867 & 0.792 & 0.979 & 0.825 & 0.844 & 0.743 & 0.997 & 0.867 & \cellcolor[rgb]{ 1,  .753,  0}0.894 & 0.628 & 0.998 & 0.812 & \cellcolor[rgb]{ 1,  .753,  0}0.892 & 0.813 & 0.979 & 0.630 & \cellcolor[rgb]{ 1,  .753,  0}0.736 & 0.532 \\
          & Mordred & 572   & 18    & 32    & 1.000 & 0.854 & 0.894 & 0.872 & 1.000 & 0.869 & 0.902 & 0.763 & 0.998 & 0.779 & 0.798 & 0.762 & 1.000 & 0.855 & 0.849 & 0.739 & 0.994 & 0.732 & 0.768 & 0.504 & 0.996 & 0.708 & 0.747 & 0.770 & 0.995 & 0.568 & 0.599 & 0.455 \\
    \bottomrule
    \end{tabular}%
}
\captionof*{table}{Table S2. Full performance metrics for best random forest classification models.}
\label{tab:full_rf_classif_metrics}%

  \thispagestyle{empty}
  \centering
  \resizebox{\linewidth}{!}{%
    \begin{tabular}{|l|lccc|cccc|rrrr|rrrr|rrrr|rrrr|rrrr|rrrr|}
    \toprule
    \multicolumn{1}{|c|}{\multirow{2}[4]{*}{\textbf{Splitter}}} & \multicolumn{1}{c}{\multirow{2}[4]{*}{\textbf{Features}}} & \multicolumn{1}{c}{\multirow{2}[4]{*}{\textbf{\# Hidden layers}}} & \multicolumn{1}{c}{\multirow{2}[4]{*}{\textbf{Hidden layer sizes}}} & \multicolumn{1}{c|}{\multirow{2}[4]{*}{\textbf{Training epochs}}} & \multicolumn{4}{c|}{\textbf{ROC AUC}} & \multicolumn{4}{c|}{\textbf{PRC AUC}} & \multicolumn{4}{c|}{\textbf{Accuracy}} & \multicolumn{4}{c|}{\textbf{Precision}} & \multicolumn{4}{c|}{\textbf{Recall}} & \multicolumn{4}{c|}{\textbf{NPV}} & \multicolumn{4}{c|}{\textbf{MCC}} \\
\cmidrule{6-33}          &       &       &       &       & \multicolumn{1}{p{4.085em}}{Training set} & \multicolumn{1}{p{4.835em}}{Validation set} & \multicolumn{1}{p{3.25em}}{Test set} & \multicolumn{1}{p{3.585em}|}{External set} & \multicolumn{1}{p{5em}}{Training set} & \multicolumn{1}{p{5em}}{Validation set} & \multicolumn{1}{p{5em}}{Test set} & \multicolumn{1}{p{5em}|}{External set} & \multicolumn{1}{p{5em}}{Training set} & \multicolumn{1}{p{5em}}{Validation set} & \multicolumn{1}{p{5em}}{Test set} & \multicolumn{1}{p{5em}|}{External set} & \multicolumn{1}{p{5em}}{Training set} & \multicolumn{1}{p{5em}}{Validation set} & \multicolumn{1}{p{5em}}{Test set} & \multicolumn{1}{p{5em}|}{External set} & \multicolumn{1}{p{5em}}{Training set} & \multicolumn{1}{p{5em}}{Validation set} & \multicolumn{1}{p{5em}}{Test set} & \multicolumn{1}{p{5em}|}{External set} & \multicolumn{1}{p{5em}}{Training set} & \multicolumn{1}{p{5em}}{Validation set} & \multicolumn{1}{p{5em}}{Test set} & \multicolumn{1}{p{5em}|}{External set} & \multicolumn{1}{p{5em}}{Training set} & \multicolumn{1}{p{5em}}{Validation set} & \multicolumn{1}{p{5em}}{Test set} & \multicolumn{1}{p{5em}|}{External set} \\
    \midrule
    Random & Mordred & 3     & 16,9,5 & 264   & 0.995 & 0.901 & \cellcolor[rgb]{ 1,  1,  0}0.912 & 0.802 & 0.992 & 0.884 & \cellcolor[rgb]{ 1,  1,  0}0.86 & 0.849 & 0.976 & 0.8   & \cellcolor[rgb]{ 1,  1,  0}0.859 & 0.767 & 0.96  & 0.763 & \cellcolor[rgb]{ 1,  1,  0}0.81 & 0.781 & 0.991 & 0.725 & \cellcolor[rgb]{ 1,  1,  0}0.85 & 0.883 & 0.992 & 0.823 & \cellcolor[rgb]{ 1,  1,  0}0.895 & 0.732 & 0.953 & 0.58  & \cellcolor[rgb]{ 1,  1,  0}0.709 & 0.478 \\
    \midrule
    Scaffold & Mordred & 3     & 16,12,3 & 359   & 1     & 0.875 & 0.742 & 0.812 & 1     & 0.866 & 0.648 & 0.869 & 0.993 & 0.75  & 0.717 & 0.743 & 1     & 0.769 & 0.689 & 0.832 & 0.985 & 0.652 & 0.689 & 0.75  & 0.988 & 0.738 & 0.741 & 0.623 & 0.987 & 0.496 & 0.43  & 0.468 \\
    \bottomrule
    \end{tabular}%
}
\captionof*{table}{Table S3. Full performance metrics for best open data based classification models.}
\label{tab:full_pub_classif_metrics}%
\end{landscape}%
\clearpage

  \thispagestyle{empty}
  \begin{landscape}
  \centering
  \resizebox{\linewidth}{!}{%
    \begin{tabular}{|llccc|ccc|ccc|rrr|}
    \toprule
    \multicolumn{1}{|c}{\multirow{2}[4]{*}{\textbf{Splitter}}} & \multicolumn{1}{c}{\multirow{2}[4]{*}{\textbf{Features}}} & \multicolumn{1}{c}{\multirow{2}[4]{*}{\textbf{\# hidden layers}}} & \multicolumn{1}{c}{\multirow{2}[4]{*}{\textbf{Hidden layer sizes}}} & \multicolumn{1}{c|}{\multirow{2}[4]{*}{\textbf{Training epochs}}} & \multicolumn{3}{c|}{\textbf{$R^2$}} & \multicolumn{3}{p{15em}|}{\textbf{Mean absolute error}} & \multicolumn{3}{p{15em}|}{\textbf{RMS error}} \\
\cmidrule{6-14}          &       &       &       &       & \multicolumn{1}{p{5em}}{Training set} & \multicolumn{1}{p{5em}}{Validation set} & \multicolumn{1}{p{5em}|}{Test set} & \multicolumn{1}{p{5em}}{Training set} & \multicolumn{1}{p{5em}}{Validation set} & \multicolumn{1}{p{5em}|}{Test set} & \multicolumn{1}{p{5em}}{Training set} & \multicolumn{1}{p{5em}}{Validation set} & \multicolumn{1}{p{5em}|}{Test set} \\
    \midrule
    \multicolumn{1}{|l}{\multirow{3}[2]{*}{AVE bias minimizing}} & ECFP  & 1     & 16    & 16    & 0.620 & 0.209 & 0.214 & 0.338 & 0.489 & 0.492 & 0.487 & 0.618 & 0.600 \\
          & MOE   & 3     & 128,32,16 & 125   & 0.841 & 0.498 & 0.368 & 0.208 & 0.342 & \cellcolor[rgb]{ 1,  .753,  0}0.346 & 0.317 & 0.507 & 0.521 \\
          & Mordred & 2     & 64,4  & 202   & 0.813 & 0.503 & 0.264 & 0.258 & 0.406 & 0.437 & 0.343 & 0.523 & 0.584 \\
    \midrule
    \multirow{3}[2]{*}{Random} & ECFP  & 1     & 16    & 14    & 0.577 & 0.351 & 0.345 & 0.360 & 0.472 & 0.466 & 0.501 & 0.618 & 0.605 \\
          & MOE   & 2     & 32,15 & 349   & 0.891 & 0.583 & 0.435 & 0.180 & 0.327 & \cellcolor[rgb]{ 1,  1,  0}0.343 & 0.259 & 0.487 & \cellcolor[rgb]{ 1,  1,  0}0.529 \\
          & Mordred & 2     & 64,12 & 478   & 0.927 & 0.601 & \cellcolor[rgb]{ 1,  .753,  0}0.557 & 0.129 & 0.296 & 0.373 & 0.201 & 0.482 & 0.577 \\
    \midrule
    \multirow{3}[2]{*}{Scaffold} & ECFP  & 3     & 16,15,13 & 229   & 0.814 & 0.360 & -0.045 & 0.263 & 0.525 & 0.641 & 0.308 & 0.687 & 0.849 \\
          & MOE   & 1     & 128   & 133   & 0.673 & 0.554 & \cellcolor[rgb]{ 1,  1,  0}0.521 & 0.260 & 0.426 & 0.422 & 0.405 & 0.598 & 0.572 \\
          & Mordred & 2     & 64,14 & 24    & 0.608 & 0.506 & 0.426 & 0.352 & 0.489 & 0.514 & 0.447 & 0.604 & 0.630 \\
    \bottomrule
    \end{tabular}%
  
}
\captionof*{table}{Table S4. Full performance metrics for best neural network regression models.}
\label{tab:full_nn_regr_metrics}%

  \thispagestyle{empty}
  \centering
  \resizebox{\linewidth}{!}{%
    \begin{tabular}{|llccc|ccc|ccc|rrr|}
    \toprule
    \multicolumn{1}{|c}{\multirow{2}[4]{*}{\textbf{Splitter}}} & \multicolumn{1}{c}{\multirow{2}[4]{*}{\textbf{Features}}} & \multirow{2}[4]{*}{\textbf{\# Trees}} & \multirow{2}[4]{*}{\textbf{Max depth}} & \multicolumn{1}{c|}{\multirow{2}[4]{*}{\textbf{Max features per split}}} & \multicolumn{3}{c|}{\textbf{$R^2$}} & \multicolumn{3}{p{15em}|}{\textbf{Mean absolute error}} & \multicolumn{3}{p{15em}|}{\textbf{RMS error}} \\
\cmidrule{6-14}          &       &       &       &       & \multicolumn{1}{p{5em}}{Training set} & \multicolumn{1}{p{5em}}{Validation set} & \multicolumn{1}{p{5em}|}{Test set} & \multicolumn{1}{p{5em}}{Training set} & \multicolumn{1}{p{5em}}{Validation set} & \multicolumn{1}{p{5em}|}{Test set} & \multicolumn{1}{p{5em}}{Training set} & \multicolumn{1}{p{5em}}{Validation set} & \multicolumn{1}{p{5em}|}{Test set} \\
    \midrule
    \multicolumn{1}{|l}{\multirow{3}[2]{*}{AVE bias minimizing}} & ECFP  & 81    & 18    & 32    & 0.871 & 0.233 & 0.184 & 0.225 & 0.476 & 0.488 & 0.284 & 0.609 & 0.612 \\
          & MOE   & 61    & 18    & 32    & 0.945 & 0.441 & \cellcolor[rgb]{ 1,  1,  0}0.405 & 0.128 & 0.379 & 0.365 & 0.187 & 0.535 & \cellcolor[rgb]{ 1,  .753,  0}0.506 \\
          & Mordred & 11    & 26    & 32    & 0.903 & 0.396 & 0.160 & 0.157 & 0.427 & 0.464 & 0.246 & 0.576 & 0.624 \\
    \midrule
    \multirow{3}[1]{*}{Random} & ECFP  & 46    & 36    & 32    & 0.906 & 0.392 & 0.324 & 0.168 & 0.444 & 0.464 & 0.236 & 0.598 & 0.615 \\
          & MOE   & 61    & 100   & 32    & 0.943 & 0.618 & 0.397 & 0.128 & 0.322 & 0.376 & 0.186 & 0.465 & 0.546 \\
          & Mordred & 107   & 13    & 32    & 0.922 & 0.559 & 0.557 & 0.151 & 0.355 & 0.429 & 0.208 & 0.507 & 0.577 \\
    \multirow{3}[1]{*}{Scaffold} & ECFP  & 20    & 100   & 32    & 0.890 & 0.339 & 0.195 & 0.160 & 0.507 & 0.558 & 0.237 & 0.698 & 0.745 \\
          & MOE   & 46    & 13    & 32    & 0.934 & 0.529 & 0.504 & 0.125 & 0.431 & \cellcolor[rgb]{ 1,  1,  0}0.420 & 0.181 & 0.615 & 0.582 \\
          & Mordred & 26    & 13    & 32    & 0.918 & 0.459 & 0.405 & 0.141 & 0.472 & 0.469 & 0.205 & 0.632 & 0.641 \\
    \bottomrule
    \end{tabular}%
}
\captionof*{table}{Table S5. Full performance metrics for best random forest regression models.}
\label{tab:full_rf_regr_metrics}%
\end{landscape}%